\documentclass{emulateapj}
\usepackage{graphicx}

\pdfoutput=1

\def\tablenotetext#1{$^{\rm #1}$}

\shorttitle{Toward a complete understanding of the MS}
\shortauthors{Hammer et al.}

\begin{document}


\title{The Magellanic Stream System. I. Ram-pressure tails and the relics of the collision between the Magellanic Clouds}


\author{F. Hammer, Y. B. Yang, H. Flores, M. Puech and S. Fouquet\altaffilmark{1}}
\affil{GEPI, Observatoire de Paris, CNRS, 5 Place Jules Janssen, Meudon F-92195, France}
\email{francois.hammer@obspm.fr}


\altaffiltext{1}{Currently at Nicolaus Copernicus Astronomical Center, Bartycka 18, PL-00-716 Warsaw, Poland}


\begin{abstract}
We have analyzed the Magellanic Stream (MS) using the deepest and the most resolved H I survey of the Southern Hemisphere (the Galactic All-Sky Survey). The overall Stream is structured into two filaments, suggesting two ram-pressure tails lagging behind the Magellanic Clouds (MCs), and resembling two close, transonic, von Karman vortex streets. 
The past motions of the Clouds appear imprinted in them, implying almost parallel initial orbits, and then a radical change after their passage near the N(H I) peak of the MS. This is consistent with a recent collision
between the MCs, $200-300$ Myr ago, which has stripped their gas further into small clouds, spreading them out along a gigantic bow shock, perpendicular to the MS. The Stream is formed by the interplay between stellar
feedback and the ram pressure exerted by hot gas in the Milky Way (MW) halo with $\rho_{hot}$= $10^{-4}$ $cm^{-3}$ at 50-70 kpc, a value necessary to explain the MS multiphase high-velocity clouds. The corresponding hydrodynamic modeling provides the currently most accurate reproduction of the whole H I Stream morphology, of its velocity, and column density profiles along $L_{MS}$. The 'ram pressure plus collision' scenario requires tidal dwarf galaxies, which are assumed to be the Cloud and dSph progenitors, to have left imprints in the MS and the Leading Arm, respectively. The simulated LMC and SMC have baryonic mass, kinematics  and proper motions consistent with observations. 
This supports a novel paradigm for the MS System, which could have its origin in material expelled toward the MW by the ancient gas-rich merger that formed M31.
\end{abstract}





\section{Introduction}
\label{intro}

The Magellanic Stream (MS) system is the second largest structure in the sky after the Milky Way (MW) disk. Since the discovery of this high-velocity, elongated H I feature  \citep{vanKuilenburg72,Wannier72}, and its identification as the MS \citep{Mathewson74a,Mathewson74b}, observations have considerably improved our knowledge of its components,  the Leading Arm (LA), Magellanic Clouds (MCs) and Bridge, and the Stream, increasing also the stretch angle up to 200$\degr$ \citep{Nidever10} and perhaps 230$\degr$ \citep{Nidever13}. The latter study shows a system apparently crossing the MW disk almost two times, at the LA and also at the MS tip, near IC 10.  The MS presents a filamentary structure \citep{Wakker01,Putman03,Nidever08} with at least two main filaments \citep{Nidever08}. Its ionized counterparts in the form of H$\alpha$ emission \citep[and references therein]{Bland-Hawthorn07} may include a significant fraction of its mass. This contrasts with the absence of a stellar counterpart, which prevents distance measurements.

Despite the long time since its discovery, and the numerous observational constraints, there is still no satisfactory physical model of the MS and the LA\footnote{Quoting \cite{Mathewson12}: "we are now no closer to reaching an understanding of the origin of the MS than when I was in the Control Room of the 210-ft Parkes Radio Telescope pumping air some 40 years ago!"}. The most popular models assume a tidal tail generated from the SMC after a recent encounter with the LMC \citep{Besla12}. They account for a first infall of the Clouds according to the recent measurements of the LMC's proper motion (PM, \citealt{Kallivayalil13}).  There are still tentative reasons for modeling the MS after a double passage of the MCs \citep{Diaz11,Diaz12}, requiring high values of the MW dark matter halo concentration \citep{Guglielmo14}. It is unclear, however, whether the MCs might have preserved their gas content after a full passage (see, e.g., \citealt{vandenBergh06}).

In fact, since the study of \cite{Sembach03}, there has been mounting evidence that the MW halo contains a significant amount of hot, highly ionized gas, which may affect the gas content of nearby galaxies such as the MCs. The observed X-ray highly ionized absorption lines show a large covering fraction (72\%) with a relatively uniform temperature ($1.8-2.4 \times 10^{6}$ K, see, e.g., \citealt{Gupta12}). Furthermore the spatial distribution of both absorption and emission is evidence of a halo origin for this hot gas, by discarding a significant contribution from the disk or from the bulge \citep{Miller13,Gupta13}. Models by \cite{Fang13} also incorporate indirect constraints such as the absence of gas in MW dSphs (from \citealt{Grcevich09}), and the pressure-confined high-velocity clouds  (HVCs, \citealt{Fox05}). In their semantic study describing the multiphase components of HVCs over the whole sky, \cite{Kalberla06} already argued that densities up to $\rho_{hot}$= $10^{-4} cm^{-3}$ are required at MS and LA locations. This is comparable to what is needed for the ionized emission of the MS itself \citep{Bland-Hawthorn07}.
Tidal scenarios (e.g., \citealt{Besla12}) for the MS have then to face considerable difficulties:
\begin{enumerate}
\item One filament of the MS likely originates from the LMC \citep{Nidever08}, a fact also supported by recent measurements of its metal abundance \citep{Richter13}, which is inconsistent with the tidal model;
\item A tidal tail becomes dominated by ram pressure for $\rho_{hot}$ $\sim$ $10^{-4}cm^{-3}$ (see \citealt{Diaz12} and Appendix A of \citealt{Besla12});
\item The much smaller size of the LMC H I disk when compared to its stellar disk is also highly suggestive of a ram-pressure truncation exerted by the halo hot gas \citep[and references therein]{Mastropietro09,Nidever14} with a density of $10^{-4} cm^{-3}$ at 50 kpc;
\item A tidal tail produced by the MC interaction invariably shows a small H I column density (one tenth of that observed, see \citealt{Besla12}); the tidal model underestimates the MS gas by almost a factor one hundred when accounting for the huge amount of ionized gas ($2\times10^{9}M_{\odot}$, \citealt{Fox14}).
\end{enumerate}

The above arguments rather support a ram-pressure model for the MS, and it has led us to reconsider the \cite{Mastropietro10} model accounting for a first passage of the MCs. The latter model described only the single LMC falling into the MW halo, while the very gas-rich SMC was ignored. Perhaps adding the SMC may take better account of the two filaments detected near the MCs by \cite{Nidever08}. However, it does not suffice to model the whole Magellanic System, since the LA cannot be plausibly reproduced by ram pressure exerted on the MCs.

The LA covers a considerable solid angle on the sky and consists of four main structures \citep{Nidever08,Venzmer12}, which makes it difficult to reproduce by a single or even a double tidal tail. Stellar outflows have been invoked by \cite{Nidever08} to explain their dynamical link with the LMC, though no modeling has been attempted to reproduce the observations. The numerous head-tails clouds populating the LA suggest they were decelerated by drag forces exerted by the halo hot gas \citep{Venzmer12}.

Besides this, ram pressure exerted on small, gas-rich dwarfs that had passed through the MW halo earlier than the LMC may have left imprints similar to the LA (see, e.g., \citealt{Yang14}). Indeed ram pressure in conjunction with tidal stripping of gas-rich dwarfs (e.g., dIrrs) may explain the formation of MW dSphs  (see review by \citealt{Mayer10}). In such a scenario, several dwarfs, including the MCs, may have reached the MW in a coordinated way, and the leading ones would have formed the LA through their deposition of gas (see Figure 11 of \citealt{Yang14}). 

The LMC and SMC are part of the Vast Polar structure of Satellites (VPOS, \citealt{Pawlowski12}) together with most MW dSphs and the MS \citep{Pawlowski13}. If the dSph progenitors have been exhausted of their gas through ram pressure following orbits unscripted in the VPOS, their imprints should be consistent with a location at the LA. The model of \cite{Yang14} explicitly assumes tidal dwarf galaxies (TDGs) coordinated along a tidal tail, and explains both the dynamical link with the LMC and the dSph kinematics. Nevertheless the possibility of a group of dwarves \citep{D'Onghia08} has also to be also investigated.

Whether the LA lis inked to stellar outflows or to trails of leading dwarfs, it left us the MS to be interpreted. 
 In this paper we examine how the MS can be described through an interaction of the Clouds with the MW halo hot gas, and to verify whether this mechanism can be at the root of the whole Magellanic System. In Section 2, we describe the main observational features, including those obtained from the most advanced survey of the MS (the Galactic All-Sky Survey, GASS, \citealt{Kalberla10}). It allows us to identify two ram-pressure tails (and then to decompose them) as well as prominent imprints of the recent MC collision, 200-300 Myr ago. It leads to a simple description of the MC orbits, which is given in Section 3, together with a scenario for the origin of the whole Magellanic System.  In Section 4, the simulated MS results from the interplay between feedback and ram pressure, and offers an order-of-magnitude gain in precision for reproducing the Stream. We also discuss the accuracy and limitations of our modeling and we solve the apparent discrepancy between the LMC orbit and the ram-pressure model, then propose a complete and novel explanation of the whole Magellanic System.

\begin{figure*}
\epsscale{1.1}
\plotone{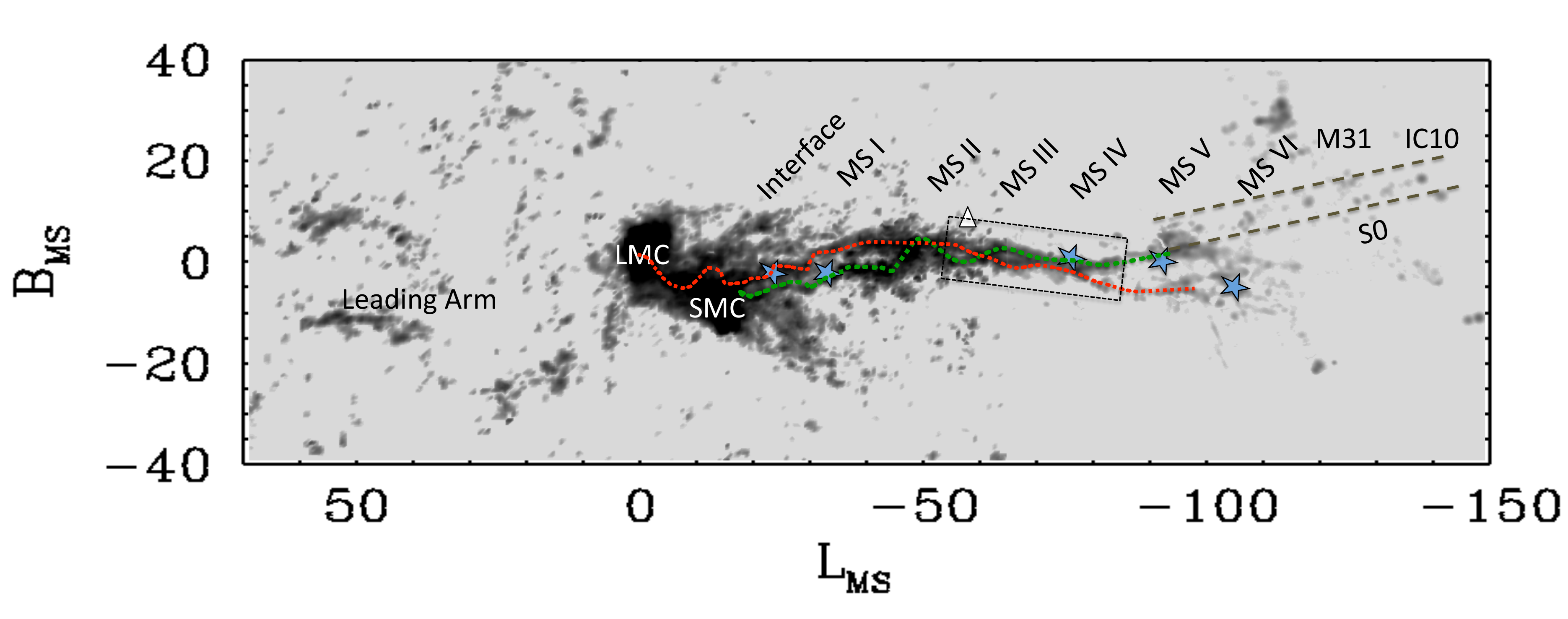}
\caption{N(H I) map  from Nidever et al. (2010) with indications of the MS components (see, e.g., \citealt{Putman03,Nidever13}) and of M31 and IC10. The five blue stars identify the five locations of active galactic nuclei observed by \citet{Fox13} and \citet{Richter13} and used for metallicity measurements. They are, from the left to the right, Fairall 9 (0.5), RBS 144 (0.075), PHL 2525 ($<$0.23), NGC 7714 (0.058) and 7469 (0.1), for which the estimated metallicities ($Z/Z_{\odot}$) are given in parenthesis. The dashed, rectangular area indicates the zoom from MS-II-IV observed by GASS that allowed us (see Figure~\ref{Fig2_MS_DNA} and Sect.~\ref{filaments}) to distinguish the two filaments shown with dotted lines (green and red, associated to the SMC and LMC, respectively), which are also based on the study of \cite{Nidever08} from the Clouds to MS-I. The open triangle marks the location of the South Galactic Pole.
}
\label{Fig1_MS}
\end{figure*}

\section{Main observational features}
\label{observations}

\subsection{Is the MS structured on two ram-pressure tails?}
\label{filaments}
 Figure~\ref{Fig1_MS} shows the H I MS from the LA to the MS tip, with the MS divided into six regions. At the Interface and MS-I regions, it reveals the two filaments identified by \cite{Nidever08} after a Gaussian decomposition of the H I velocity profile. The filamentary structure is even more prominent from MS-II to MS-IV, and quoting \cite{Putman03}, the two filaments of the Stream give the impression of twisting about each other in a double helix, akin to a DNA molecule. At the locations MS-V and MS-VI the structure is clearly more complex, with the continuation of the MS presenting four filaments \citep{Stanimirovic08}, as well as many structures in a wide area including M31. The additional, prominent S0 filament found by \cite{Nidever13} is indicated in Figure~\ref{Fig1_MS}. 
 
\subsubsection{Nearly parallel filaments from MS-II to MS-IV}

\begin{figure}
\epsscale{1.1}
\plotone{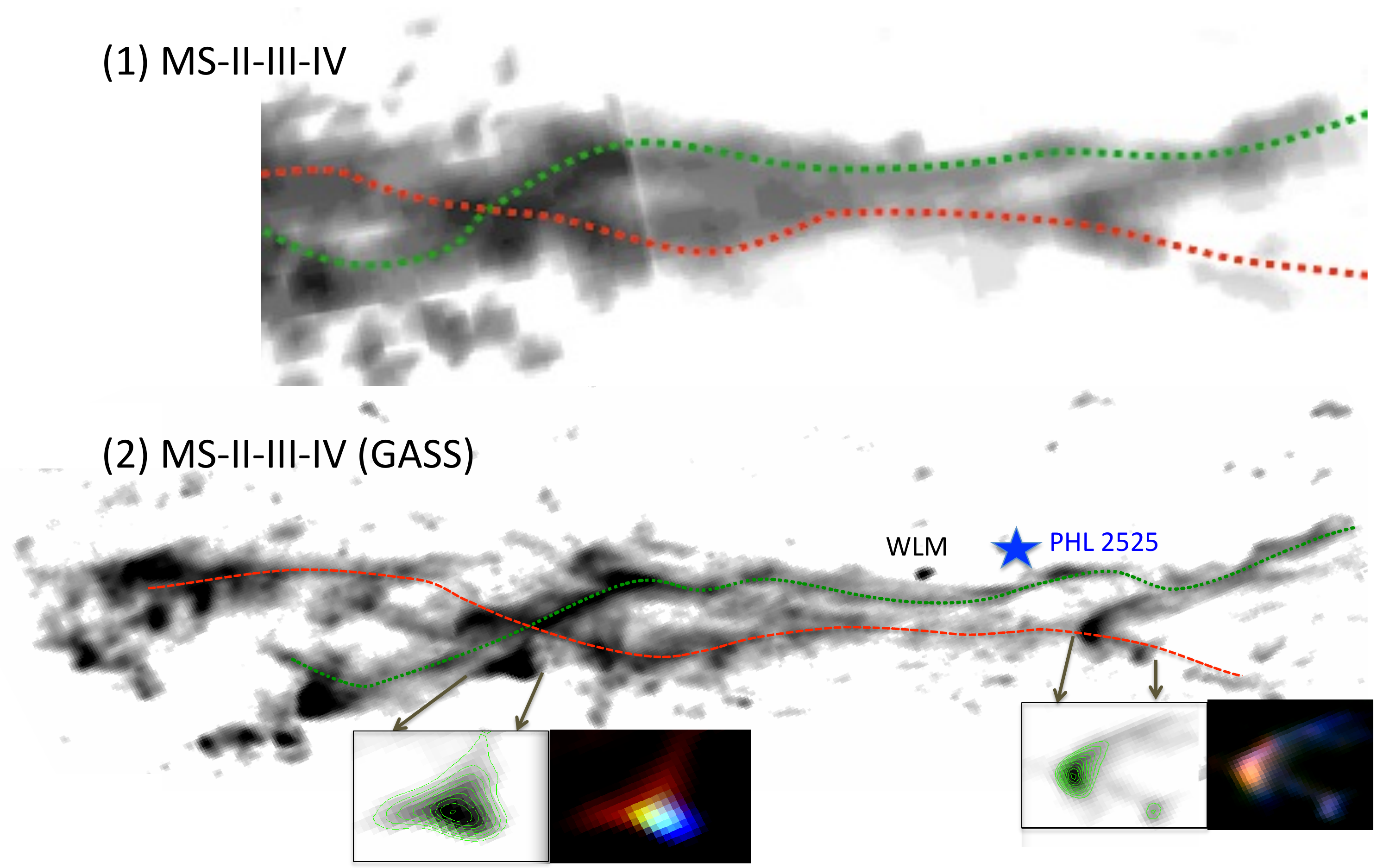}
\caption{ (1) Zoom of Figure~\ref{Fig1_MS} in the area from MS-II to MS-IV. (2) Column-density map of the Magellanic Stream integrated from $V_{LSR}$= -40 to -400 km$s^{-1}$ after using the GASS data cube. Red (green) dotted lines indicate the filaments coming from the LMC (SMC), and the WLM dIrr location is indicated as well as the background active galactic nucleus, PHL 2525.  A zoom of a bright clump associated with the LMC filament is shown, with on the left, a RGB color image coded from B= -225 $\pm$ 5 km$s^{-1}$ and R= -201.5 $\pm$ 5 km$s^{-1}$. A zoom of a bright clump associated to the SMC filament is shown with, on the left, an RGB color-coded image from B= -127.7 $\pm$ 3.2 km$s^{-1}$ and R= -108.0 $\pm$ 3.2 km$s^{-1}$. Red (blue) color is assumed to represent largest (smallest) velocity values. }
\label{Fig2_MS_DNA}
\end{figure}

The spatial resolution of Figure~\ref{Fig1_MS} was not sufficient to recover the numerous details of the filaments from MS-II to MS-IV. We have retrieved the 
 data from the Parkes GASS \citep{McClure09}, which is the highest spatial resolution (14 arcmin), 
 the most sensitive (1$\sigma$$T_B$ noise 57mK) at 1 km$s^{-1}$ resolution, and currently the most accurate H I survey of the Southern Hemisphere  \citep{Kalberla10}. Figure~\ref{Fig2_MS_DNA} shows a zoom of the stream at the locations MS-II, III and IV and provides evidence of these improvements when comparing Panels 1 and 2. By examining the data cube, we have identified the locations where the filaments are crossing each other and  hence recovered the twisting filamentary system. Figure~\ref{Fig2_MS_DNA} illustrates a tentative decomposition of the MS into the two individual filaments. Such a decomposition is quite trivial for most of the length of the MS, except for a region near WLM where the filaments are more diffuse. Assuming the above, the overall filaments can be followed in Figure~\ref{Fig1_MS}, from the MS tip to the MCs. 

\begin{figure}
\epsscale{1.10}
\plotone{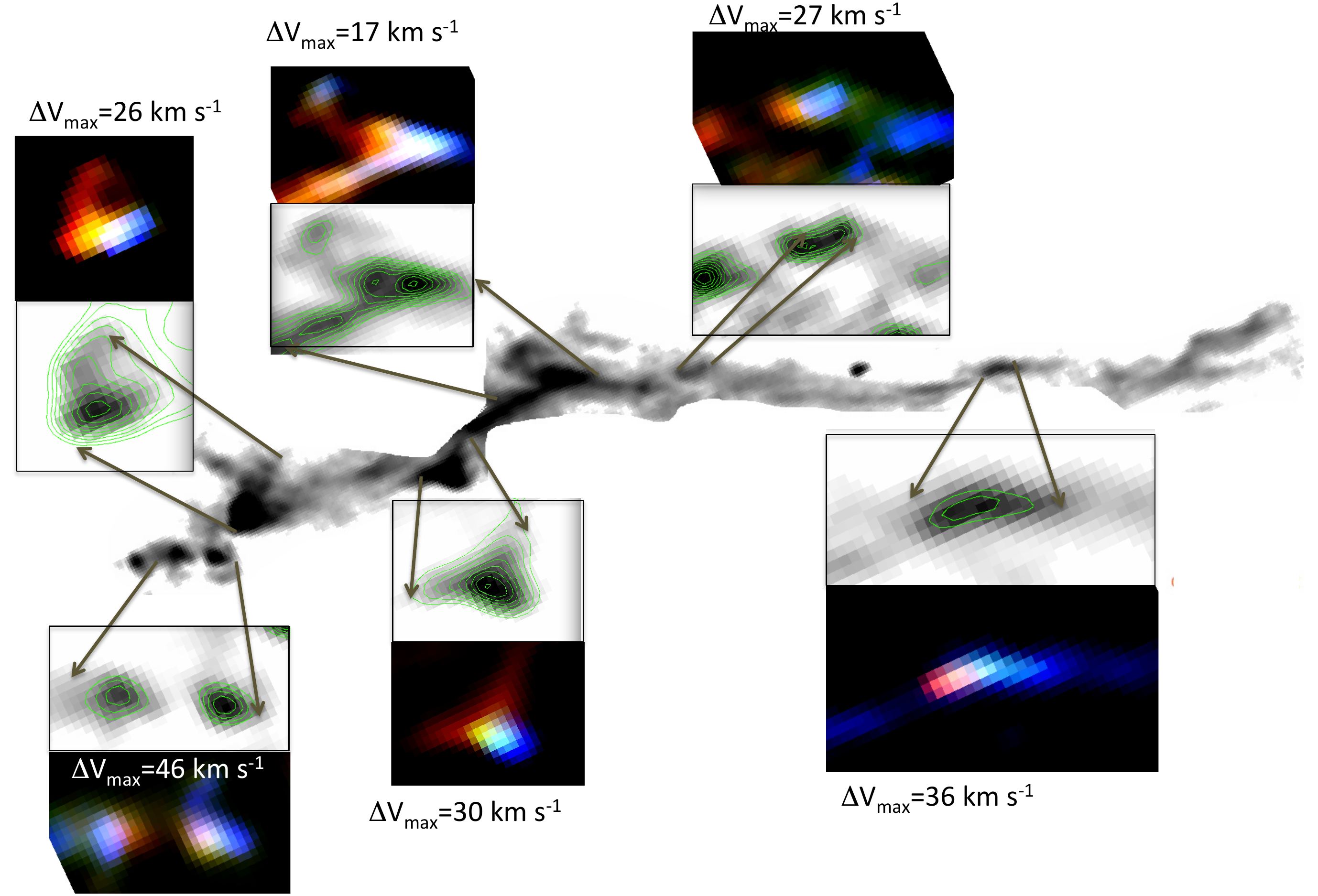}
\caption{The SMC filament and its associated vortices. For each vortex insets show a zoom with contoured lines and an RGB color image, for which each color is made of the sum of from 9 to 21 channel of the original GASS data cubes. Red (blue) color is assumed to represent largest (smallest) velocity values. Along the filament, from left to right, vortex $(l, b)$ coordinates in degrees are: (348.43,-79.87), (352.25, -81.14), (31.01, -81.43), (60.26, -81.53), (70.55, -78.16), and (78.22, -69.83). }
\label{Fig3_Vortices}
\end{figure}

There are only a few examples of observed H I ram-pressure tails, such as the one provoked by the infall of NGC 4388 in the very hot intracluster medium (ICM) of Virgo \citep{Oosterloo05}. The corresponding 120 kpc ram-pressure tail shows an oscillatory (or S) shape, a property shared by the two filaments associated with the MS (Figure~\ref{Fig2_MS_DNA}). The development of KelvinÐHelmholtz (KH) instabilities depends on the Reynolds number above critical values studied by \cite{Roediger13}. We have used their Equation (1) assuming a Spitzer temperature-dependent viscosity, and following \cite{Bland-Hawthorn07}, we adopt\footnote{Here we use V $\sim$ 300 km$s^{-1}$ instead of 350 km$s^{-1}$ \citep{Bland-Hawthorn07} on the basis of our modeling (Section \ref{modeling}) from MS-II to MS-IV where we also measure the characteristic widths of the filaments to be $\lambda$= 2$\pm$ 0.25 kpc,  assuming a distance of 70 kpc.} V=300 km$s^{-1}$, $\lambda$= 2 kpc, $n_{e}$= 2 $10^{-4}$ $cm^{-3}$, and T= 1.5 $10^{6}$ K for the velocity, characteristic size of the filaments, and for the density and temperature of the hot medium, respectively. It results that
\begin{equation}
 Re= 445 \times f_{\mu}^{-1},
\end{equation}

in which the viscosity suppression factor $f_{\mu}$ $\le$ 1 depends on the role of the magnetic field and on the importance of various physical mechanisms such as cooling (see, e.g., \citealt{Roediger13}). Because the MW halo gas is colder than the Virgo ICM, the MS is likely more turbulent than the tail of NGC 4388. \cite{Roediger13} showed that critical Reynolds values are in the range of a few tens to one hundred, for a density contrast ($\rho_{cold}/\rho_{hot}$) of 100 \citep{Bland-Hawthorn07}, thus confirming the turbulent behavior of the MS filaments.

In the following we aim at verifying whether the filamentary structure of the MS can be interpreted as the result of ram-pressure tails trailing behind the MCs.

\subsubsection{The MS: two turbulent, transonic von Karman vortex streets}
\label{vonKarman}
 The characteristic time for the KH instabilities to grow is (see, e.g., \citealt{Bland-Hawthorn07}):
\begin{equation}
\label{tKH}
(\tau_{KH}/1Myr)= 6.52  (\rho_{cold}/\rho_{hot})^{0.5}  (\lambda/2kpc)
 (V/300 kms^{-1})^{-1}
\end{equation}

The orbital time (2$\pi$$D/V$) is
\begin{equation}
 (\tau_{orb}/1Myr)= 1027 \times (D/50kpc) \times  (V/300 kms^{-1})^{-1}
\end{equation}

Assuming that most of the MS is composed of two filaments related to each MC implies that a large extent of the MS is dominated by KH instabilities. This is plausible since for a reasonable density contrast of $\rho_{cold}/\rho_{hot}$ from 100 to 1000, $\tau_{KH}$ is smaller than  $\tau_{orb}$ for cloud lengths up to $\lambda$= 10 kpc.  
 We have tested whether or not the prominent clumps in the filaments can be related to the formation of vortices. In fact similar structures are quite often prominent in simulations, e.g., see Figure 6 of \cite{Stone97}, or in nature\footnote{See e.g., http://landsat.visibleearth.nasa.gov/view.php?id=625 revealing a unique pattern called a (laminar) Òvon Karman vortex streetÓ expected in a subsonic regime, while \cite{Stone97} studied another kind of regular pattern in a supersonic regime.}. However, the Stream is a transonic flow \citep{Bland-Hawthorn07} with a Mach number of $\sim$ 1.5. For Mach numbers around one, a number of different shedding configurations appear in addition to the traditional von Karman vortex street behavior \citep{Gostelow05,Gostelow12}. Moreover, for $Re$ values ranging from 300 to 3 $\times 10^{5}$, the vortex street is expected to be turbulent (e.g., \citealt{Lienhard66}). Vortices are expected to be condensed and rotating structures, being at the tip of one or more sub-filaments. We identified nine of them in each of the two MC filaments, which are bright clumps at the edge of one or two sub-filaments, and elongated as it could be expected for an inclined rotating disk. Figure~\ref{Fig3_Vortices} illustrates how the resolving power of GASS allows us to scrutinize the sub-components of the MS flow. Exploring the GASS data cube, we have been able to spatially and kinematically resolve all the nine elongated clumps of the SMC filament, finding a clear and strong rotation well aligned to the major axis in all of them, and with full velocity gradients ranging from 17 to 46 km$s^{-1}$ (see insets in Figure~\ref{Fig3_Vortices}). We estimate the velocity dispersion of these structures to be $\sim$10 km$s^{-1}$. This is evidence of the presence of rotationally supported structures as expected in most vortices. One case (see the second top panel from the left) is slightly more complex as it is made of three vortices very close to one another, though rotation is clearly dominating them.

The LMC filament also shows an oscillatory spatial behavior with clumps that resemble vortices, as illustrated in Figure~\ref{Fig2_MS_DNA} (see the zoom of a vortex candidate in panel 2). However, these vortex candidates appear to be different than those linked to the SMC filament. The velocity gradient along their elongated shapes is either not resolved (see Figure~\ref{Fig2_MS_DNA}) or it is due to a mixture of several unresolved peaks. We have not been able to resolve any of them. If corresponding bright clumps in the LMC filament are vortices, they should not be spatially resolved by GASS. This is quite a reasonable hypothesis, because the vortices associated with the SMC filament are resolved by merely $3-4$ resolution elements. How has such a striking difference between the two filaments been generated? They should have both experienced a similar ram pressure and KH instabilities given that they lie close to each other. The only difference is their metallicity content. Metallicity strongly affects the cooling and we conjecture that this could explain the different behaviors of the two filaments from MS-II to MS-IV.

The finding that each filament possesses different types of vortices strengthens our initial attempt at disentangling the MS into two filaments or ram-pressure tails all the way from the Clouds to MS-IV (see Figure~\ref{Fig1_MS}). We also retrieved the GASS data in the Interface and MS-I region, confirming the analysis by \cite{Nidever08} of a MS dominated by two filaments, one reaching the LMC and the second with a more uncertain origin, either the SMC or the Bridge; the considerable number of streams and filaments prevents us from improving the  conclusions of \cite{Nidever08}. 

\subsubsection{Ram pressure tails and Stream abundances}
Figure \ref{Fig1_MS} shows the five   active galactic nuclei that lie in the direction of the Stream. The difference in metallicity between the first two (Fairall 9 and RBS 144, \citealt{Fox13} and \citealt{Richter13}),  in the Interface and MS-I regions, brings a striking confirmation of the dual origin for the MS, with one filament associated with the LMC as identified by \cite{Nidever08} and the other with the SMC. The remaining three lines of sight lie in the Stream tip end and indicates relatively low metallicities which, at first glimpse, is expected for the SMC rather than for the LMC material. \\

\cite{Fox13} cautioned against a too simplistic an interpretation of their data when comparing gas-phase abundances in the Stream with present-day abundances in the
Clouds (LMC: $Z/Z_{\odot}$=0.46; SMC: $Z/Z_{\odot}$=0.22). For example, in a ram-pressure model, the infalling Cloud progenitors are gradually stripped from their outer to their inner gas disk during their journey to the pericenter. Then one expects that metallicity toward the Stream tip increasingly reflects the metallicity of the outer regions of its parent galaxy at the time at which it was stripped, $\sim$ $1-2$ Gyr ago, and not its present value. Figure 9 of \cite{Meschin14} reveals that some $1-2$ Gyr old RGB stars at $5\degr.5-7\degr.1$ from the center of the LMC may have metal abundances similar to those of NGC 7469 ($Z/Z_{\odot}$=0.1). At the very tip of the MS, the latter (near the potential LMC filament) has almost twice the metal abundant of NGC 7714, which would be near the potential SMC filament (see Figure \ref{Fig1_MS}).
Within uncertainties, the location and abundances of the five QSOs on the Stream seem to be quite consistent with two ram-pressure tails as they are observed (see Figures \ref{Fig1_MS} and \ref{Fig2_MS_DNA}). \\

\subsection{Residuals of a close interaction between MCs}
\label{residue}

\subsubsection{Observational evidence for a recent past collision}
There is compelling evidence that the LMC has had a recent and close encounter with the SMC. 
 First, the Magellanic Bridge is a tidal feature that is likely reproduced by a recent encounter (see, e.g., \citealt{Besla12}). Second, their PMs imply at least one collision within the past 500 Myr, as a direct result of the relative orientation of their three-dimensional velocity vectors \citep{Kallivayalil13}. Third, the distribution of OB stars in the MCs and near the Bridge is also consistent with a recent exchange of material, $\sim$ 200 Myr ago \citep{Casetti13}.

\begin{figure}
\epsscale{1.10}
\plotone{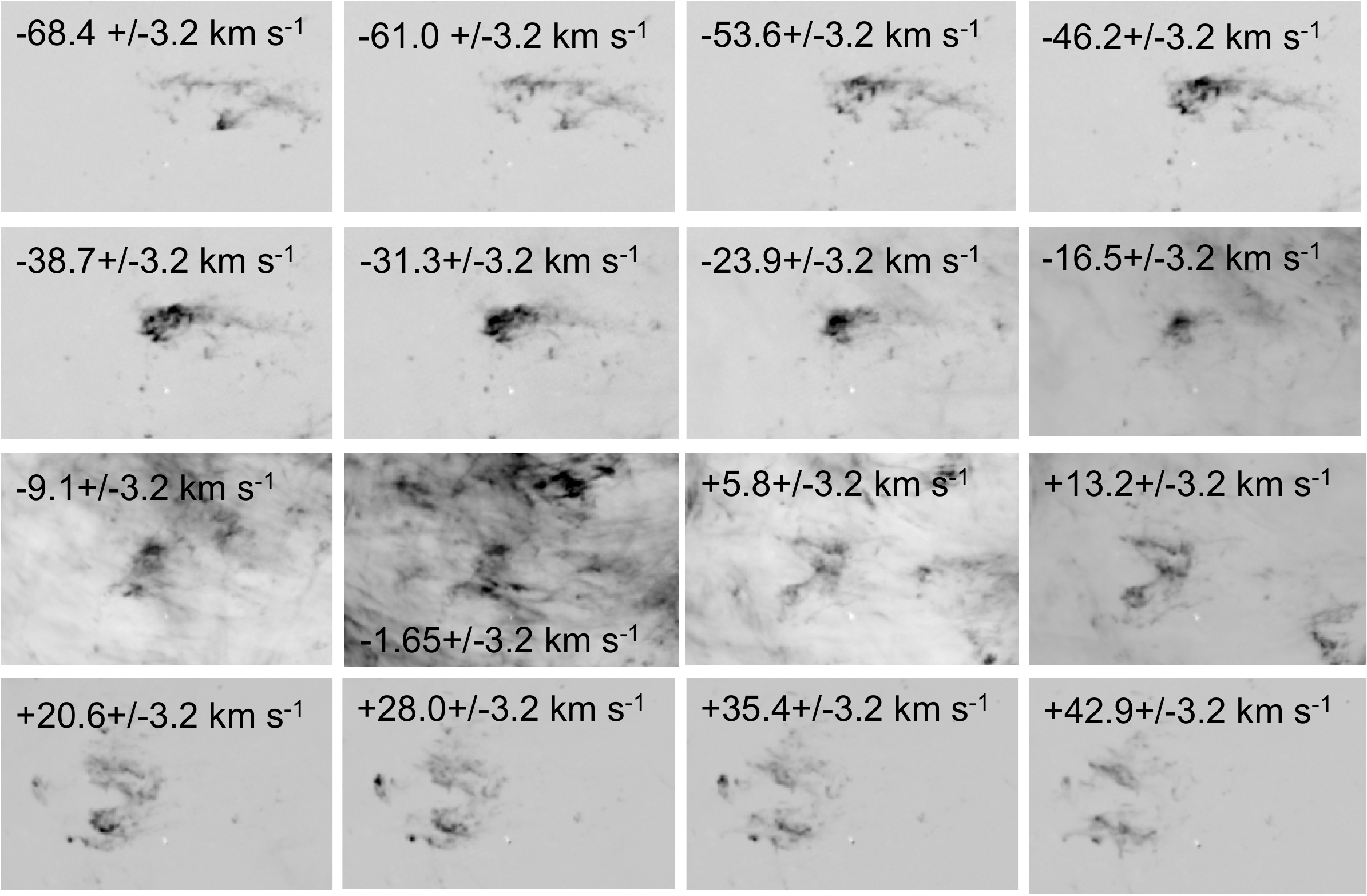}
\caption{Magellanic Stream at the South Galactic Pole from -68.4 to +42.9 km $s^{-1}$. Each stamp represents 39 $\times$ 24.5 $deg^{2}$. The exquisite spectral resolution of the GASS survey is unique to the study of the MS very near Galactic emission. Each image is a combination of 9 velocity channels of GASS, using the IRAF "imcombine" subroutine and applying a 3$\sigma$ clipping algorithm to eliminate any spurious and low-dispersion signal. This series of images can be used to identify what is linked to the MS. Galactic emissions vary very rapidly from one image to another, in contrast to the MS whose structure is far more stable.}
\label{Fig4_collision}
\end{figure}

\begin{figure}
\epsscale{1.20}
\plotone{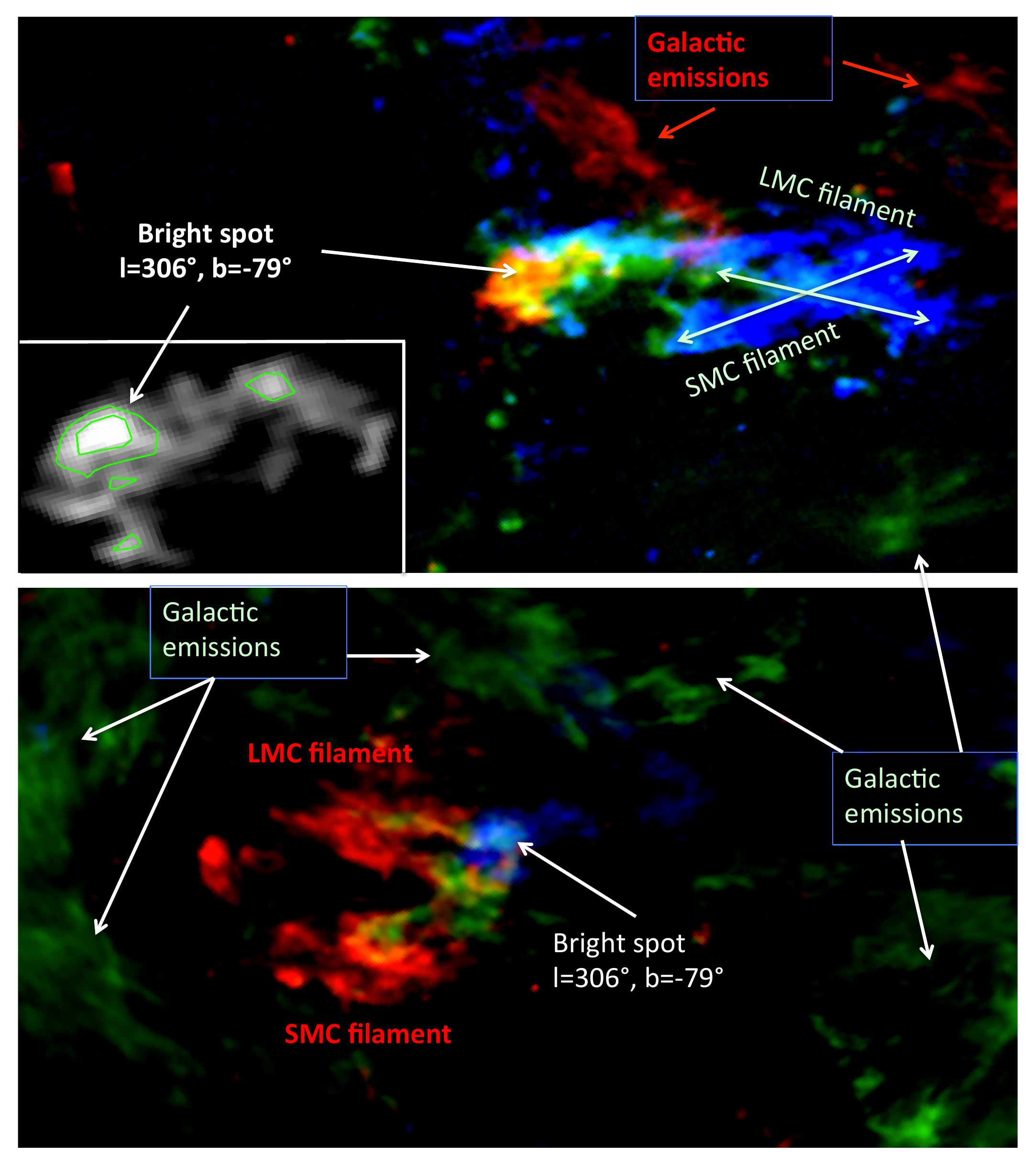}
\caption{Behavior of the Magellanic Stream when reaching the South Galactic Pole; each panel is 39 $\times$ 24.5 $deg^{2}$. Each color image has been obtained by summing up 36 original GASS velocity channels. They are aligned and centered on the bright spot at $l= 306\degr$ and $b= -79\degr$ (red and blue in the top and bottom panels, respectively). {\it Top:} RGB color map with R= -23.5 $\pm$ 14.4 km $s^{-1}$, G= -53.1 $\pm$ 14.4 km $s^{-1}$ and B= -82.8 $\pm$ 14.4 km $s^{-1}$. The inset zooms the sum of the three maps at the surroundings of the bright spot for N(H I)$>$ 8 $\times 10^{20}$ $cm^{-2}$, the green contours are limited by 1.7 and 2.3 $10^{21}$ $cm^{-2}$. Galactic emissions (mostly in red) are indicated. {\it Bottom:}  R= +36 $\pm$ 14.4 km $s^{-1}$, G= +6.3 $\pm$ 14.4 km $s^{-1}$ and B= -23.5 $\pm$ 14.4 km $s^{-1}$. Galactic emissions  (mostly in green) are indicated. }
\label{Fig5_SGP}
\end{figure}

\subsubsection{The unusual properties of the MS near the South Galactic Pole (SGP)}
\label{SGP}
In the context of a ram-pressure scenario there should be signatures or imprints of a gigantic collision in the recent past. To identify them in the GASS data cube may be straightforward because velocity is a rough tracer of the epoch at which MCs were depositing gas material through ram-pressure stripping\footnote{Deposited gas clouds are lagging behind the MCs because they are decelerated by ram pressure exerted by the hot gas.}. Figure~\ref{Fig1_MS} indicates that filaments are crossing for the last time between MS-I and MS-II before reaching the MCs. This happens near the SGP, which is strongly contaminated by the MW emission. Figure~\ref{Fig4_collision} shows the MS within 16 velocity slices, each summing nine channels of the original GASS data cube to improve the signal-to-noise ratio (S/N) and the contrast with emission from the MW. Velocity steps are sufficiently narrow to disentangle most of the MS from the various MW emissions, especially those that are spurious with small velocity dispersions. From -31.3 to -9.1 km $s^{-1}$ the MS emission is concentrated on a very bright spot at $l=  306\degr$ and $b= -79\degr$ ($L_{MS}$= -47.5$\degr$ and $b_{MS}$= 2.3$\degr$). Within this velocity range and a diameter of 80$\degr$, there is no such bright emission elsewhere. Then a contamination by the MW emission is very unlikely. Figure~\ref{Fig5_SGP} shows the extent of the MS on both sides of the bright spot. The latter shows no evidence for a velocity gradient and appears to be a complex mixture of knots and filaments (see the zoomed view in Figure~\ref{Fig5_SGP}). It comprises the only velocity range for which the MS does not show the double filament structure.

\begin{figure*}
\epsscale{1.}
\plotone{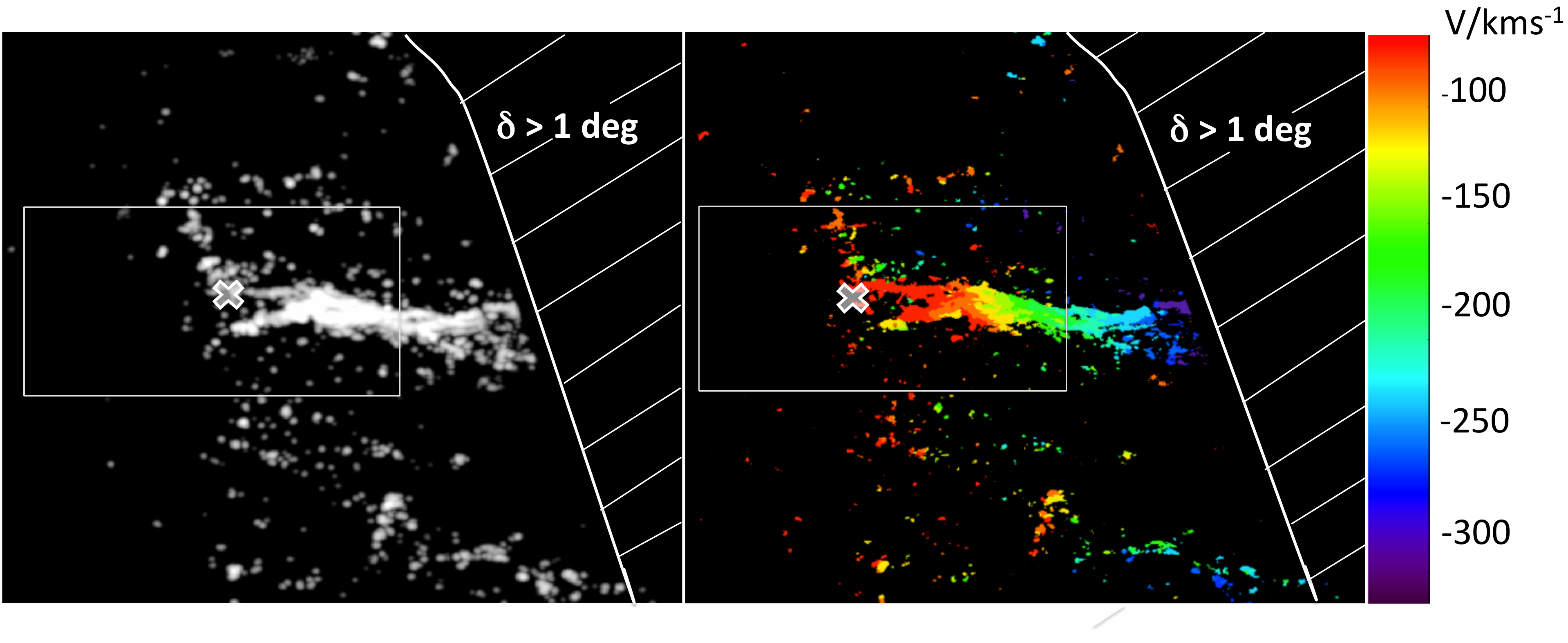}
\caption{Behavior of the Magellanic Stream when reaching the SGP.  In both panels, the rectangular box identifies the location and size of the two panels in  Figure~\ref{Fig5_SGP} and the cross represents the location of the bright spot. {\it Left:} sum of the GASS velocity slices within -90 to -334 km$s^{-1}$. Negative and low S/N values (S/N$<$ 2) have been set to zero before the summation to improve the final S/N. The line and the hatched area indicate the limits of the GASS survey ($\delta$ $\le$ 1 $\degr$). {\it Right:} GASS velocity slices from -90 to -334 km$s^{-1}$ that show the MS and the numerous surrounding HVCs with color-coded velocities. Notice the concentration of HVCs on a front line almost perpendicular to the MS and passing through the cross. }
\label{Fig6_shock}
\end{figure*}

The unusual properties observed near the SGP include the following:
\begin{itemize}
\item with a peak at N(H I)= 3.5 $\times 10^{21}$ $cm^{-2}$, the bright spot is the highest peak of the MS (see also Figure 10 of \citealt{Nidever10});
\item it coincides with a strong excess of ionized, warm gas material  \citep{Madsen12,Bland-Hawthorn13};
\item at the bright spot $L_{MS}$ (-47$\degr$.5) there is a significant excess of HVCs with anomalous velocities  showing a considerable velocity scatter from -250 to 250 km$s^{-1}$ (\citealt{Putman02,Putman03} and Figure~\ref{Fig6_shock}).
\end{itemize}

The above properties suggest the residual near the SGP of a quite energetic event. The excess of anomalous HVCs could not be associated with the background Sculptor Group as convincingly shown by \cite{Putman03}. The origin of the ionized gas peak has been proposed by \cite{Bland-Hawthorn13} to be caused by a powerful flare at the Galactic Center $2-3$ Myr ago, though such a mechanism can be responsible neither for the high density of H I gas at the bright spot nor for the anomalous kinematics of associated HVCs. 

The  bright spot and its immediate surroundings in Figure~\ref{Fig5_SGP} (see the inset) account for almost 10\% of the whole Stream N(H I). We suspect that this gas has been stripped by the collision of the LMC with the SMC, during which the gaseous disk of the SMC progenitor  would have been fully destroyed (see, e.g., \citealt{Nidever14}). The very peculiar morphology of the bright spot may suggest that it could have been further shrunk by the ram pressure during its travel, lagging behind the MCs. At a distance of $\sim$ 70 kpc, a velocity of $\sim$ 300 km $s^{-1}$, and a size of $\sim$ 2 kpc, Eq. \ref{tKH} shows that it can survive for more than $t_{KH}$= 450 Myr, i.e., more than the time elapsed since the collision.

\subsubsection{ Relics of gas after the collision and residuals of a gigantic bow shock, pressurized by the halo hot gas}
\label{shock_relics}
Figure~\ref{Fig6_shock} reveals the residuals of a gigantic bow shock extending over several tens of degrees on both sides of the bright spot (see the cross symbol). It comprises numerous small clouds with a large scatter of velocities, revealing in large number the anomalous HVCs of \cite{Putman02,Putman03}. The formation of small H I cloud debris is not unexpected in gas-rich galaxy collisions, though because of their small size they are rapidly heated and destroyed by the hot gas through KH instabilities (see Eq. \ref{tKH}). They concentrate toward a front line (see the left part of Figure~\ref{Fig6_shock}) almost perpendicular to the MS and passing through the bright spot (see the cross). We verified the GASS data cube, and while confirming even the presence of HVCs with positive velocities on the right of this front line, we have been unable to detect more than a few tiny HVCs on the left of this line. 
We suggest that anomalous HVCs are associated with a recent collision between the MCs, which has expelled large numbers of small clouds into the hot medium. Their absence beyond the front line can be attributed to their fast ionization when they entered to an increasingly hot gas density, from the right to the left, at increasing $L_{MS}$. 

\section{The ram-pressure plus collision scenario: toward a complete understanding of the Magellanic System}
\subsection{Consequences for the MC orbits}

\begin{figure*}
\epsscale{1.1}
\plotone{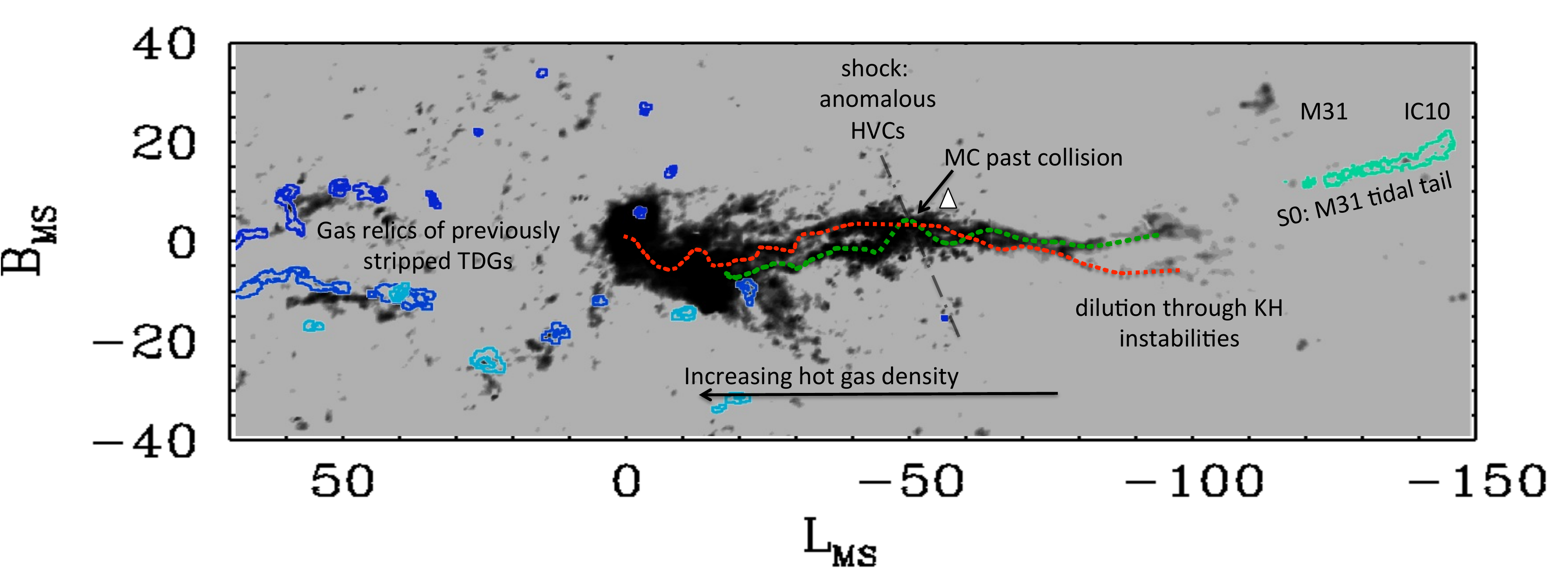}
\caption{N(H I) map  from Nidever et al. (2010) on which the proposed scenario is superimposed.   The imprint of the first M31 merger tidal tail (see \citealt{Fouquet12,Hammer13}) is predicted to coincide in location, shape, and velocity with the S0 filament that extents toward IC10. Here we assume that the M31 tidal tails cannot survive within the 200 kpc (300 kpc) halo hot gas of the MW (M31). Cloud progenitor orbits pass through or close to S0, and then are rather well traced by the filaments (dotted lines, red: LMC, green: SMC). The Leading Arm (LA) is reproduced by gas deposited by previously stripped dwarf galaxies (based on improved models of \citealt{Yang14}) and the different blue colors distinguish each individual passage. The open triangle marks the location of the SGP.
}
\label{Fig7_MS}
\end{figure*}

\cite{Kalberla06} examined the physics of HVCs with $\sigma$ $\sim$ 12 km $s^{-1}$ near the MS and they discovered a surprising multiphase medium that was not expected at such distances. 
\label{Orbits}
Examination of their Figures 19 and 20 shows the huge increase of multiphase HVCs from the MS tip end to the MCs. 
  This may indicate a higher confining halo gas pressure and consequently a decreasing distance of the MS, from the tip to the MCs as indicated by the horizontal arrow in Figure~\ref{Fig7_MS}. 
This is expected if the MS is the product of ram-pressure tails generated by the stripping of the MC gas: in such a case, the MCs could have just entered the halo hot gas near the MS tip end ($L_{MS} \sim$ -130$\degr$), and have had more and more gas stripped by the denser hot gas with increasing $L_{MS}$. 
 The findings of numerous head-tail HVCs in the MS and in the LA regions support the need for a ram-pressure stripping mechanism \citep{For14}.

 As shown in Sect.~\ref{filaments} (see also \citealt{For14}) the MS mostly consists of two ram-pressure tails, which implies that from MS-II to MS-IV, the MCs were following rather parallel orbits along the observed MS. The oscillatory behavior and then the DNA-like appearance could be attributed to the superposition of two transonic, von Karman vortex streets (see Figure~\ref{Fig2_MS_DNA}). The above orbital conditions have been radically changed from MS-I to the MCs (see Figures~\ref{Fig7_MS} and \ref{Fig5_SGP}).  As shown in Sect.~\ref{residue} a collision may well explain this, assuming that the SMC had been kicked out after a close encounter with the LMC. 
 In the top panel of Figure~\ref{Fig5_SGP}, one may recognize both the LMC and SMC filaments as shown in  Figure~\ref{Fig2_MS_DNA}. In the bottom panel the two filaments show a much wider angle (70$\degr$-80$\degr$), which appears as a direct imprint of a radical change of trajectory, especially for the SMC. Figure \ref{Fig7_MS} summarizes the ram-pressure plus collision scenario for the MS.

\subsection{A Ram-pressure origin for the LA} 
 
 It is currently argued (see, e.g., \citealt{Besla12}) that a purely ram-pressure solution cannot pull material forward to explain the LA feature.
Alternatively, the LA could have been formed from gas deposited by MW dSph progenitors having been gas-stripped by the MW hot gas, as modeled by \cite{Yang14}. They have been able to reproduce most of their properties, including their kinematics. They assumed their progenitors to be TDGs coming at large velocities from a tidal tail formed at the  location of M31. Large velocities help the gas-rich TDGs to be fully stripped by the halo hot gas after a single passage, some of them also being captured by the MW (see \citealt{Fouquet12}). In  Figure 11 of \cite{Yang14}, the gas residuals after the passage of gas-stripped progenitors of dSphs show striking similarities with the LA, both in space, shape, velocity, and column density. We report the same mechanism in Figure~\ref{Fig7_MS} but accounting for three leading passages of TDGs. 
 
It can be argued that one cannot find enough dSphs with recent star formation, which would be implicitly assumed to reveal a recent gas stripping, a few Gyr ago. Only Fornax, Leo I and Carina show recent star formation \cite[see their Figure 11]{Weisz14}, and Fornax should furthermore be excluded because it has not even reached its pericenter \citep{Yang14}. However, the need for recent star formation is relevant for large dIrrs such as LMC/SMC, but not for the far much smaller progenitors of the present-day MW dSphs. In \cite{Yang14} (see their Figure 9 for their lowest mass TDG) it was verified that the gas rapidly decouples from stars and then reaches regions that are close to the LA, explicitly because they followed orbits similar to the MCs. Then star formation may have stopped on entering the halo $2-3$ Gyr ago. It may even happen that the gas density could have not or hardly reached the Kennicutt-Schmidt limit, implying an absence of new stars after the formation of small, very gas-rich TDGs, i.e., almost 9 Gyr ago, reconciling the scenario with many small dSph progenitors.

\subsection{A novel paradigm for the Origin of the Magellanic System}
\label{MS_origins}
 The Magellanic System could be caused by the successive passages of leading, small dwarfs that have lost their gas from ram pressure followed by the far more massive gas-rich MC progenitors losing their gas in the form of the MS. Such a mechanism is plausible since the orbits of the MW dSph and MC progenitors are all inscribed into the vast polar structure (VPOS) surrounding the MW  \citep{Kroupa05,Pawlowski11,Pawlowski12}. It has been suggested that these progenitors may have entered the MW halo as a group \citep{D'Onghia08}. However, the large time difference for MC and dSph progenitors to form respectively the MS and the LA for entering the MW halo would invariably leave the LMC progenitor behind by $\sim$ 100 kpc. Since the latter stellar mass is the larger, by factors ranging from 25 to 100, a group is to be excluded.

The MS extends from the Clouds to the M31 system, especially toward IC10 through the S0 filament discovered by \cite{Nidever13}. M31 is almost certainly the result of an ancient gas-rich merger (\citealt{vandenBergh05}, pp. $1-15$; \citealt{Courteau11,Kormendy13}) due to its light distribution including a nearly classical bulge and its metal rich halo. This is indeed the case for many spiral galaxies that are thought to have had their disks rebuilt  through gas-rich major mergers \citep{Hammer05,Hammer09,Stewart09,Hopkins10a,Hopkins10b} during the past $\sim$ 10 Gyr. The formation of M31 has been modeled by \cite{Hammer10} with a 3:1 gas-rich polar interaction. With a first passage and fusion at 8.75$\pm$ 0.35 and 5.5$\pm$ 0.5 Gyr ago, respectively, it can reproduce simultaneously the properties of the Giant Stream, the 10 kpc ring, the thick disk, and most of the internal structure of M31. The fusion epoch also seems consistent with the recently established age-dispersion relation for the M31 disk \citep{Dorman15}.

Although constructed in 2010, the M31 modeling also predicts \citep{Hammer13} the co-rotating plane of satellites discovered by \cite{Ibata13}. The latter includes the tidal tail formed after the first passage of the M31 progenitors, which is predicted to lie in the direction of the MW, consistently with a plausible trajectory for the MC and MW dSph progenitors to form the VPOS \citep{Yang10,Fouquet12}. The M31 co-rotating plane of satellites is indeed pointing toward the MW within 1$\degr$ \citep{Ibata13}.

The S0 filament lies all the way from the Stream tip to IC 10. IC 10 lies in the co-rotating plane of satellites surrounding M31 \citep{Ibata13} and it is precisely within the tidal tail expelled during the first passage of the M31 merger, 8.5 Gyr ago (compare Figure~\ref{Fig7_MS} with the Figure 2 of \citealt{Hammer13}). Projected on the sky, this tidal tail should appear as an almost linear feature. Figure~\ref{Fig7_MS} illustrates how the projection of the M31 tidal tail is similar to the S0 filament in space. In our modeling, the S0 filament indicates the path along which TDGs are expected to reach the MW halo (see Figure~\ref{Fig7_MS}).

Fouquet et al. (2015) study the morphological and stellar population of similar systems of TDGs expelled up to 8.5-9 Gyr ago. They are fed by the tidal tail, are stable systems not fully supported by rotation, and show many similarities with dIrrs. Studies of TDGs have been done mostly in the neighborhood of their host galaxy (see also \citealt{Ploeckinger14,Ploeckinger15a,Ploeckinger15b} for studies of their stability), which includes the discovery of a $\sim$ 4 Gyr old one by \cite{Duc14}.

One may object, however, that it leads to too exceptional a Local Group if, among its two main galaxies, one (M31) is an ancient merger literally sending to the other (the MW) a tidal tail emitted at the first passage of the interaction. For a polar interaction, merger theory predicts that the first tidal tail is particularly well aligned with the plane of the orbital angular momentum, the latter also dominating the re-formed disk. We do observe the M31 almost edge-on, and this observation makes an interaction of the tidal tail with the MW much more likely.  This has led us \citep{Hammer13} to claim an intimate link between the M31's plane of satellites and the vast polar structure (VPOS) surrounding the MW  \citep{Pawlowski11}. This also elegantly solves the issue of the proximity of the Clouds to the MW (see, e.g., \citealt{vandenBergh10}), a property shared by only 0.4\% of similar galaxies \citep{Robotham12}.

 Figure~\ref{Fig7_MS} illustrates that the Stream could be the most striking prediction of such a link, as it delineates a 150$\degr$ line on the sky from the MCs to the M31 system, and more precisely toward IC 10. It is the goal of this paper to verify the plausibility of the new paradigm\footnote{Since a merger at the location of M31 predicts the disks of satellites surrounding both the MW and M31 (see \citealt{Hammer13}), one does not need to account for the {\it 'look-elsewhere'} effect to interpret their statistical relevance as proposed by \cite{Cautun15}.}

\section{Modeling the MS and MCs}
\label{modeling}

\subsection{Hydrodynamical model of the ram-pressure plus collision scenario}
\subsubsection{Goals and limitations of the simulation}
Table 1  summarizes the properties that have to be reproduced by our modeling and evaluates  the success of previous models (tidal: \citealt{Besla12} and \citealt{Diaz12}; ram-pressure: \citealt{Mastropietro10}) or of a not modeled yet scenario (outflow : \citealt{Nidever08}). It illustrates that no existing models can reproduce all or even a significant part of the MS properties. Table 1 also suggests that both ram pressure and MC interaction could be necessary to recover the MS properties: for example, the Bridge evidences a recent tidal interaction between the MCs. Outflows could be also necessary to deliver enriched gas at the outskirts.\\

\begin{table*}
 \centering
 \begin{minipage}{15cm}
  \caption{Properties of the Magellanic System to be Reproduced and Performances of Three Families of Scenarios}
 \label{scenario}
 \begin{tabular}{lllll}
  \hline
  \hline
  &   & Scenarios  & &\\  
 \hline
    &   & Ram & Stellar & References\\
           Property & Tidal & Pressure & Outflows & Notes\\
 \hline
 \hline
 Streams:        &  &  &  & \\
 \hline
No stars & N & Y & Y & \\
 Length     & Y & Y & N & 240$\degr$, \cite{Nidever13}\\
 Orientation       &  Y & Y & N & toward M31\\
N(H I) density        & N & Y\tablenotemark{a} & Y & \cite{Nidever10}\\
Leading Arm       & Y/N\tablenotemark{b} & N & Y & \cite{Venzmer12}\\
Bridge       & Y & N & N & \\
Interface, MS-I     & Y/N\tablenotemark{c} & N & N & two filaments, \cite{Nidever08}\\
MS-II,III,IV      & Y/N\tablenotemark{c} & N & N & DNA-like filaments \citep{Putman03}\\
MS-V,VI      & - &  - & - & bifurcation \citep{Stanimirovic08}\\
S0 filament & N & N & N & \cite{Nidever13}\\
  \hline
\hline
MW:      &  &  &  & \\
 \hline
Halo hot gas     & N\tablenotemark{d} & Y & - & \\
 \hline
 \hline
MCs:     &  &  &  & \\
 \hline
LMC H I disk     & N & Y & N & shrunk, \cite{Nidever14}\\
MC motions     & Y & N &  - &  \citep{Kallivayalil06,Kallivayalil13}\\
Stellar mass & Y & Y & Y &\\
Gas mass  & Y & Y & Y &\\
\hline
\end{tabular}\\
\tablenotetext{a}{\cite{Mastropietro05} used and reproduced the low N(H I) values available at this time}
\tablenotetext{b}{Tidal model never succeeds in reproducing the four Leading Arm structures found by \cite{Venzmer12}}
\tablenotetext{c}{The double filamentary structure has been reproduced after two passages \citep{Diaz12}}
\tablenotetext{d}{The presence of halo hot gas is a serious show-stopper for tidal models except if $n_h$ $<<$   $10^{-4}cm^{-3}$}
\end{minipage}
\end{table*}

Our main goal is to verify whether or not the intertwisting filamentary structure and overall morphology of the MS (see Figure~\ref{Fig7_MS}), its N(H I) and velocity profiles along $L_{MS}$, can be reproduced by two ram-pressure tails, after assuming orbits for the Clouds following a path near the S0 filament when entering the MW halo (parallel then colliding near the SGP) as it is described in Sect. \ref{Orbits}. The ram-pressure plus collision scenario requires us to verify whether TDGs can be progenitors of the MCs, including their baryonic content (stars and gas, see Sect.~\ref{IC_MC}, ~\ref{MCmasses}) and kinematics (see Sect.~\ref{LMCrot}), and their
 past orbits that should be consistent with the PM measurements of \cite{Kallivayalil13} (see Sect.~\ref{PMs} and the Appendix).

Besides the above, our analysis focuses solely on the reproduction of the Stream, i.e., without regard to the detailed morphology of the MCs such as the LMC bar, or their precise orientation on the sky for orbital calculation reasons (see Sect.~\ref{calcul_orbit}). 

We use GADGET2\footnote{The softening scale and viscosity used in each simulation are 0.05 kpc and 0.8 (the GADGET2 default value), respectively.} \citep{Springel05} with gas cooling, feedback and star formation as developed by \citet{Wang2012} (see their Appendix), in order to be consistent with our simulations of TDGs \citep{Hammer10,Fouquet12,Yang14}.  Smoothed particle hydrodynamics (SPH) codes have difficulties in reproducing KH instabilities, swamping the low-amplitude modes. For example, the mixing between stripped material and the halo hot gas is then underestimated (see, e.g., Figure 25 of \citealt{Hopkins14}). In Sect. \ref{KH_role} we verify whether or not this affects the survivability of the filamentary structure of the Stream,  and we show in Sect.~\ref{ionization} how the use of moderate resolution is further limiting our ability to reproduce the gigantic shock described in Sect. \ref{shock_relics} (see also Figure \ref{Fig6_shock}). We use a kernel of 50 to improve convergence in simulation results without being affected by the clumping instabilities (see \citealt{Valdarnini12,Hopkins14}).  
\begin{figure}
\epsscale{1.1}
\plotone{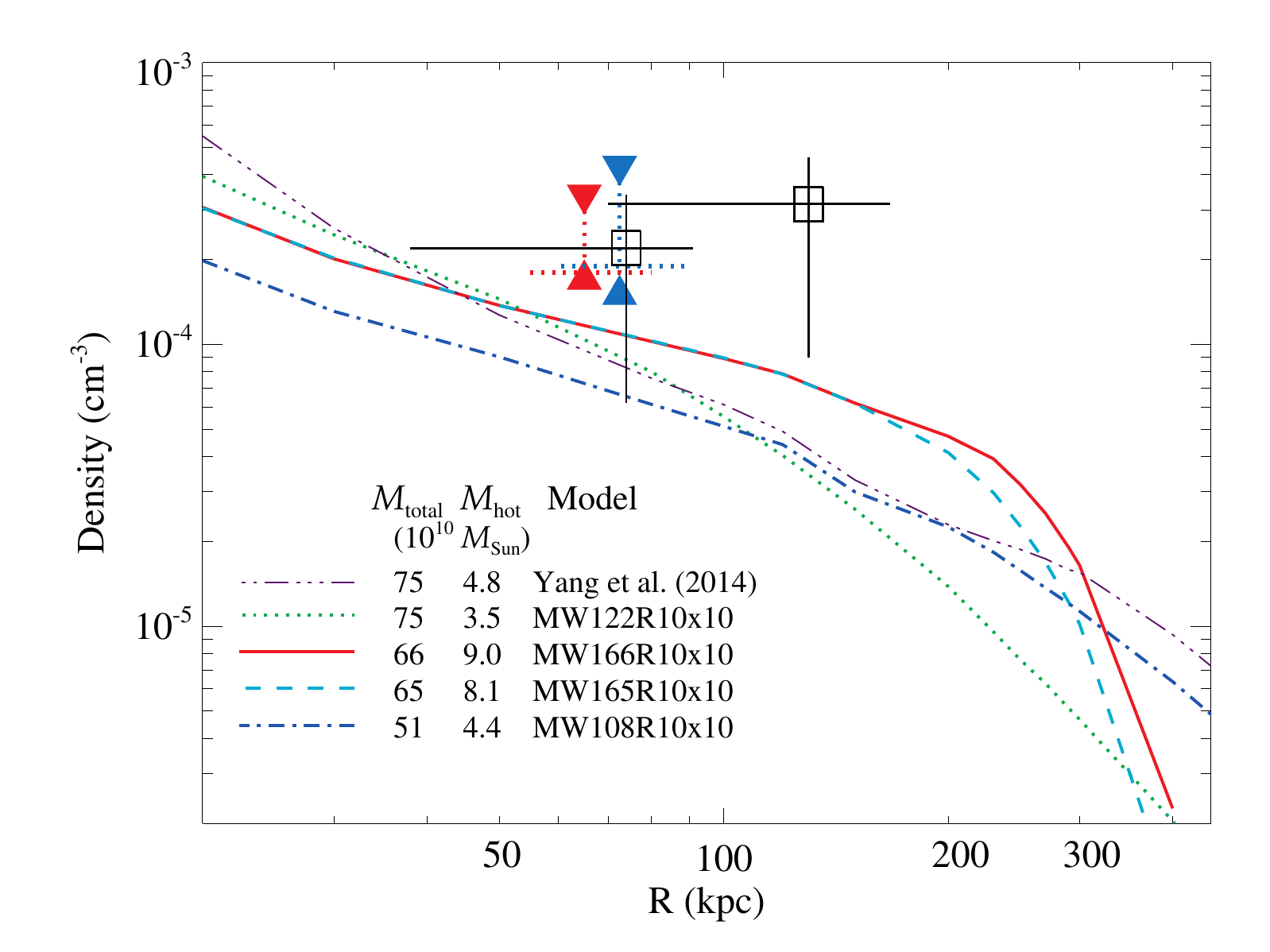}
\caption{A selection of models of the  density profile of the halo hot gas, which have been used in this paper to generate the MS after MC interaction. Total mass and hot gas mass are given in the bottom-left. Constraints due to stripped dSphs are provided: open black squares indicate those from the study of \citet{Grcevich09}, and red and blue points show those from \citet{Gatto13} (triangles represent upper and lower limits and horizontal lines correspond to the different distributions of orbits that have been tested).}
\label{Fig8_hotgas}
\end{figure}

\subsubsection{Initial conditions of the MW and of its halo hot gas}

Following \cite{Yang14}, we have constructed a MW model with four components, including a dark matter halo, a gaseous halo,
a bulge, and a disk (star+gas). We have used the same values for the bulge and disk and the same methodology to relax all these components together to warrant their
equilibrium and to match the observed MW quantities (see Table 2). The main difference\footnote{We also use a resolution ten times higher than \cite{Yang14} for the hot gas particles with a mass of 2.75 $\times 10^{5}$ $M_{\odot}$. We have verified that using the same mass for hot gas and cold gas particles left the results unchanged.} with \cite{Yang14} is that we have
left variable the hot gas profile, its mass (from 1 to 9 $\times 10^{10}$$M_{\odot}$), and also the dark matter mass (see Figure~\ref{Fig8_hotgas}). 
For the latter we assumed baryonic fractions from 0.14 to 0.19, i.e., consistent with the universal value from \cite{Komatsu11}. 
This can be verified by adding the values of the hot gas mass shown in Figure~\ref{Fig8_hotgas} to the stellar mass (5 $\times 10^{10}$ $M_{\odot}$, see \citealt{Hammer07})
and to the disk gas mass assumed to be 0.68 $\times 10^{10}$ $M_{\odot}$. 

 We first tuned up the ram pressure by using different hot gas profiles (see Table \ref{tbmw} and Figure \ref{Fig8_hotgas}). Allowing some freedom on the hot gas profile helps to reproduce the
MS, including its length and the velocity profile along the MS longitude ($L_{MS}$), which could be tuned since the gas in the tails is slowed down by ram pressure. To initiate our modeling, we have first constructed the MW $N$-body models (see Table \ref{tbmw}) and let them run in isolation for 4 Gyr to verify their stability. Indeed they become stable in less than 0.5 Gyr and remain so for 4 Gyr, to within less than 5\% for their density, temperature, and gas distribution, out to R = 500 kpc. To run our MS simulations, we export a snapshot of the isolated run of the MW model at 1 Gyr after its beginning. This ensures that our MW models are stable, because simulations have been run for a duration of 2-4 Gyr with Cloud progenitors sent at 450 or at 850 kpc from the MW. 

\subsubsection{Initial conditions for the Cloud progenitors}
\label{IC_MC}
TDGs are pre-selected from former GADGET2 realizations of a tidal tail from the M31 merger model \citep{Hammer10,Hammer13,Fouquet12,Fouquet15,Yang14}. This is one of the main limitations of our study since we have only one TDG massive enough ($M_{baryon}$= 3.2 $\times 10^{9}$ $M_{\odot}$ including 0.5 $\times 10^{9}$ $M_{\odot}$ in gas, TDG8b in \citealt{Fouquet15}, see their Table 4) in our database, which is hardly sufficient to be the LMC progenitor. We also noticed that this TDG is too compact when compared to the LMC ($r_{half}$ $\sim$ 0.8 kpc instead of $\sim$ 2.2 kpc) and not gas-rich enough to account for both the gas in the present-day LMC and that released to the MS. 

This is likely due the strongly decreasing (by a factor 5) feedback prescription adopted by \cite{Hammer10} to optimize the modeling of the M31 disk. In order to obtain a more extended TDG, we go back to the original M31 simulation and extract a cube of $120^3$ $kpc^3$ centered at the location of the TDG in the tidal tail and let this cube evolve in isolation using conditions of constant and high feedback conditions (precisely five times the median feedback value described in \citealt{Cox06} and the initial value taken by \citealt{Hammer10}) to warrant a more extended and gas-rich object. In principle this is similar to having performed the overall modeling with a constant\footnote{We also investigate a modest decrease of the feedback from 5 to 4 times the  median value of \cite{Cox06} (see Sect.~\ref{ionization}).} and high feedback. Within such conditions we obtain an initial TDG-A1 (and as well TDG-B1 for the SMC progenitor that is extracted from the same tidal tail; TDG3b in \citealt{Fouquet15}) that is gas-rich and compact enough (see Table \ref{tdg}). We have used two slightly different models of MW hot gas for TDG-A1 and B1, which are named Model 385 (with MW model 165) and Model 395 (with MW model 166) to test the interplay between feedback and ram pressure at the outskirts of the MW.  

We have also tried to account for the huge ionized gas mass ($\sim$ 2 $\times 10^{9}$ $M_{\odot}$) discovered by \citet{Fox14}, implying that, if originating from the Clouds, more than two thirds of their initial gas has been extracted from their progenitors. We estimate that an initial mass of 4.7 $\times 10^{9}$ $M_{\odot}$ could be necessary to account for gas loss by ram pressure and star formation. This leads us to generate TDG-A2 and B2 and Model 266 (see Tables \ref{tbmw} and \ref{tdg}), for which we have rescaled the TDG masses by $\sim$ 1.6 and have checked their internal equilibrium by evolving them in isolation for 3 Gyr, which is a long enough duration for our MS simulations.  

\begin{figure}
\epsscale{1.1}
\plotone{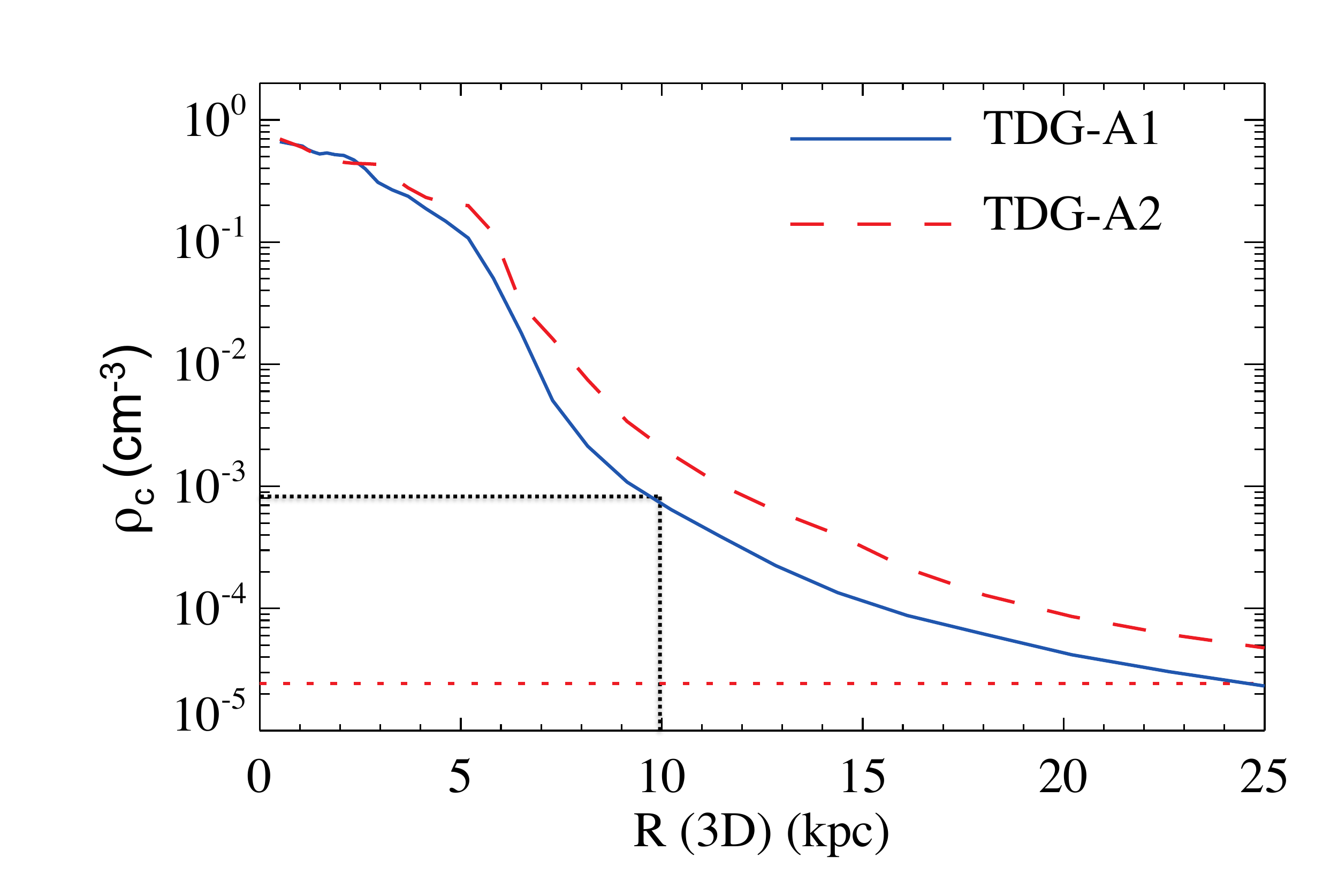}
\caption{Gas density profile of the adopted TDG for the LMC progenitor and of the associated tidal tail. The horizontal dashed line indicates the average density of the tidal tail.
}
\label{Fig9_TDG}
\end{figure}

Figure \ref{Fig9_TDG} presents the gas density profiles of TDG-A1 and A2 before their extraction from the tidal tail. During the MS simulations, initial TDGs have been extracted within 10 kpc (see dotted lines) to avoid too much undesirable low gas density that would not be correctly treated by GADGET2. We verified that our simulations are unchanged by varying this cut from 7 to 20 kpc, which we attribute to the adopted strong feedback that dominates the gas profile in the outskirts.

\begin{table*}
{\centering
\caption{Parameters of the MW model.}
{\small
\begin{tabular}{lccccccccrr} \hline \hline
Model No. & MW Model
& $m_{\rm total}$
& $m^{\rm dark}_{\rm halo}$
& $m^{\rm hot\,\,gas}_{\rm halo}$
& $B/T$\tablenotemark{a}
& $r^{\rm  stellar}_{\rm disk}$ & $r^{\rm  gaseous}_{\rm
  disk}$ &  $f_{\rm gas}$ \\
&  ($10^{10}M_{\odot}$)  & ($10^{10}M_{\odot}$)
& ($10^{10}M_{\odot}$)
& ($10^{10}M_{\odot}$)
&  & (kpc)
& (kpc) & \\ \hline
No. 266 & MW122   initial values & 77.3& 68.4&  3.4& 0.00&  0.0&  5.0& 0.20\\
 &             mean values & 74.9& 65.9&  3.5& 0.22&  2.4&  4.7& 0.13\\ \hline
No. 385 & MW165   mean values & 64.5& 50.9&  8.1& 0.16&  2.2&  4.1& 0.13\\ \hline
No. 395 & MW166   mean values & 65.6& 51.1&  9.0& 0.16&  2.2&  4.2& 0.13\\ \hline
& MW108   mean values & 50.6& 40.7&  4.4& 0.18&  2.2&  4.0& 0.16\\  \hline
\end{tabular} \\
\label{tbmw}
}
}
\tablecomments{In order to construct a galaxy model that represents the basic properties of the MW, we have artificially input a bulge-less disk with large gas extent (see, e.g., MW122). For all models we have kept the disk baryonic content to 5.5 $\times 10^{10}$ $M_{\odot}$. All quantities here are evaluated within a sphere with a radius of 260~kpc and all models have been using the same mass per particles (2.75 $\times 10^{5}$ $M_{\odot}$ for halo hot gas, 2.75 $\times 10^{6}$ $M_{\odot}$ for stars, and 1.1 $\times 10^{7}$ $M_{\odot}$ for dark matter).}
\tablenotetext{a}{$B/T$: bulge-to-total mass ratio, defined as $m_{\rm bulge}/m_{\rm star}$}
\end{table*}

\begin{table*}
{\centering
\caption{Initial and final properties of simulated MC progenitors.}
\label{tdg}
{\small
\begin{tabular}{llcccccccc} \hline \hline
 TDG  & Epoch & $m_{\rm total}$ &
  $m_{\rm gas}$  & $r_{\rm gas}^{\rm cutoff}$\tablenotemark{a} 
&$m_{\rm star}$ & $r_{\rm star}^{\rm half}$\tablenotemark{b}
& $f_{\rm gas}$\tablenotemark{c} \\
Name & (Model \#) & ($10^9M_{\odot}$) &($10^9M_{\odot}$) & (kpc)  & ($10^9M_{\odot}$)  & (kpc)  &  \\ 
(MC) & & {\small $<\!\!3.5$($<\!\!10$)kpc} &  {\small $<\!\!3.5$($<\!\!10$)kpc}  &  & {\small $<\!\!8$($<\!\!10$)kpc}  & & {\small $<\!\!3.5$($<\!\!10$)kpc}  \\
\hline   
TDG-A1    & Initial           &    1.86 (3.06)    & 1.10 (2.29) &  5.18   & 0.76 (0.77) &   1.86    &     0.59 (0.75) \\
(for LMC) & Final (395)  &    1.82 (2.99)    & 0.72 (1.78) &  1.65   & 1.10 (1.21) &   1.71      &      0.39 (0.60)  \\ 
(for LMC) & Final (385)  &     1.82 (2.99)  & 0.65 (1.78) &   4.4   & 1.18 (1.21) &  1.65  & 0.35 (0.60) \\  
                   \hline 
TDG-B1    & Initial            & 1.39 (2.24)       & 1.01 (1.84) &  4.5   & 0.38 (0.39) &   2.11      & 0.73 (0.82) \\ 
(for SMC) & Final (395)  & 1.00 (1.93)       & 0.48 (1.34) & ---   & 0.52 (0.59) &  3.63   &  0.48 (0.69) \\ 
(for SMC) & Final (385)  & 1.00 (1.83)       & 0.48 (1.28) & ---   & 0.52 (0.55) & 3.62 & 0.48 (0.70) \\ 
      \hline
\hline 
TDG-A2    & Initial       & 3.15 (4.73) & 1.36 (2.90) &   5.0 & 1.79 (1.83) & 1.66 & 0.43 (0.61)\\ 
(for LMC) & Final (266) & 2.83 (4.11) & 0.49 (1.65) &   3.8  & 2.34 (2.46) & 1.27  & 0.17 (0.40) \\
                   \hline 
TDG-B2    & Initial        & 2.35 (2.99) & 0.94 (1.53) &   3.7  & 1.41 (1.46) & 1.73  & 0.40 (0.51) \\ 
(for SMC) & Final (266) & 1.55 (2.47) & 0.30 (1.09) &   ---  & 1.25 (1.38) & 2.04 & 0.20 (0.44) \\ 
         \hline
         \hline
               \end{tabular}        \\
\tablecomments{The particle (gas and stars) masses are 2.75 $\times 10^4M_{\odot}$ for TDG-A1,B1 and 4.4 $\times 10^4M_{\odot}$ for TDG-A2,B2. Initial conditions have been set at -4 Gyr (at 850 kpc from the MW) for TDG-A1, B1 and at -2 Gyr (450 kpc) for TDG-A2, B2, in order to account for the extent of their associated MW halos (see Figure~\ref{Fig8_hotgas}). Masses for "Final" results are measured in two ways: (1) in order to compare with the observations, $m_{\rm star}$ and $m_{\rm gas}$ are measured within radii of 8 kpc and 3.5 kpc, respectively, and (2)  we also measured the masses in a sphere of 10-kpc radius (values in parenthesis).}
\tablenotetext{a}{Cut-off radius of TDG-A1 and A2 gas disks (TDG-B1, B2 gas disks are generally destroyed)}
\tablenotetext{b}{Half-mass radius of stars} 
\tablenotetext{c}{Gas fraction}
} 
}
\end{table*}

\subsubsection{Calculating Orbits of the $MCs-MW$ Interactions}
\label{calcul_orbit}
To perform a simulation of the Cloud interactions with the MW halo, we assume the scenario described in Sect. \ref{Orbits}, 
in which MCs have had close and almost parallel motions in the tidal tail, a few billion years ago, and a recent interaction that is
 responsible for their large difference in present-day velocities (LMC and SMC with $V_{tot}$= 321 and 217 km$s^{-1}$, \citealt{Kallivayalil13}). MC progenitors are assumed to be TDGs lying in a single
tidal tail, with a velocity difference of less than 30 km$s^{-1}$. Their motions have to be consistent with that of the largest tidal tail 
induced by the M31 merger, after the first passage (see, e.g., \citealt{Yang10};\citealt{Hammer10}).
\cite{Fouquet12} have shown that such a trajectory is consistent with the estimate of the M31 PM by \cite{vanderMarel12}.

We first attempt to calculate the initial positions and velocities of the Clouds, several Gyr ago (see Table~\ref{tdg}), to launch an $N$-body/SPH simulation. To do so, we used first the same $N$-body solver that is presented in \cite{Yang10}.
However, the hydrodynamics effects appear to be quite dominant when gas-rich galaxies are interacting together as well as with a hot medium. 
In order to have a bridge formed that looks similar to the one observed, we find that the collision should happen about $200-300$ Myr ago with a pericenter of $2.5-4$ kpc.  Mass and energy losses due to the {\em hydrodynamic} collision at such a small pericenter will considerably affect their dynamical state after the collision. 
One of us (YY) has improved the orbit calculation by including an empirical "hydrodynamic solver" that takes into account the energy and mass losses during the interaction between the TDGs. The hydrodynamic corrections have been studied using several TDG-pair collisions. Additionally, we also account for the slow-down of TDGs due to ram pressure of the hot medium using the formulas presented in \cite{Sofue1993}.

 This simulation of the MS formation involves 18 free parameters (2$\times$2 for MW total and hot gas mass and radius and 14 for MC positions, velocities, and spins, see below), whose optimization helps to reproduce the MS properties. Our modeling is slightly simpler than the complete optimization made by \cite{Guglielmo14}, who used genetic algorithms. This is because the masses of LMC and SMC are taken within discrete values (see Sect.~\ref{IC_MC}), and then the mass, radius, and gas fraction of the progenitors of MCs could be considered as fixed parameters for which only small variations can be adopted from the fiducial models, for example using variations of the feedback regulation. 

The total MW mass is in the lower range of current estimates (i.e., $(5-12) \times 10^{11}$$M_{\odot}$), since the scenario proposed in Sect.~\ref{MS_origins} excludes that MW dSphs could be used as mass tracers for the MW.
The mass profile of dark matter has been modeled using a core model \citep{Barnes02}, which leads to two parameters, mass and radius. Constraints on the total mass are given by the spatial location and by the kinematics of the MS that fix the orbit of the MC system in a ram-pressure model. Together with the MW rotation curves and the PM of the LMC alone, this suffices to fix the total mass profile within a dozen of simulations.
Then for given MC progenitors, we may constrain the  density distribution of MW hot gas, by visually comparing the ram-pressure tail lengths and H I column density with the observations. Several tens of simulations are necessary to optimise the two parameters ($m^{\rm hot\,\,gas}_{\rm halo}$ within $(1-10) \times 10^{10}$$M_{\odot}$ and radius within $60-500$ kpc) characterizing the MW hot gas. 

Last, we enter a fine-tuning process, in which we have reduced the number of free parameters to 14, i.e., the 3$\times$2 positions and the 3$\times$2 velocities of the LMC/SMC progenitors plus their spins. The key information that we are searching by tuning these free parameters is the collision between the MCs, i.e., the pericenter and the orbital angular momentum of the collision, which are strongly constrained by the current positions and motions of both LMC and SMC. For each simulation we have to consider the constraints provided by the H I maps of the MS by Nidever, including N(H I) intensity, kinematis projection in the ($V_{\rm LSR}$,$L_{MS}$) frame, together with the Cloud motions and locations. 
Hundreds of simulations were carried out in order to visually match the essential properties of the MS and LMC.

The above considerations allow us to predict, several Gyr ago, the locations and velocities of the MCs within an accuracy of 5\%. Better reproduction of the collision between the Clouds requires a precision of less than 1\%, which could be simply inferred from the ratio between the pericenter and the initial distance of TDGs to the MW, i.e., about 400-500 kpc, 2 Gyr ago (or 800-900 kpc, 4 Gyr ago for TDG-A1,B1). A more detailed description of the calculations will be presented in Yang et al. (2016, in preparation). In such a large parameter space and with the accuracy that we have achieved, our goal is only to look for a solution illustrating the global mechanism and physical processes derived from the analysis of the MS in Sect. \ref{observations}.

\begin{figure}
\epsscale{1.3}
\plotone{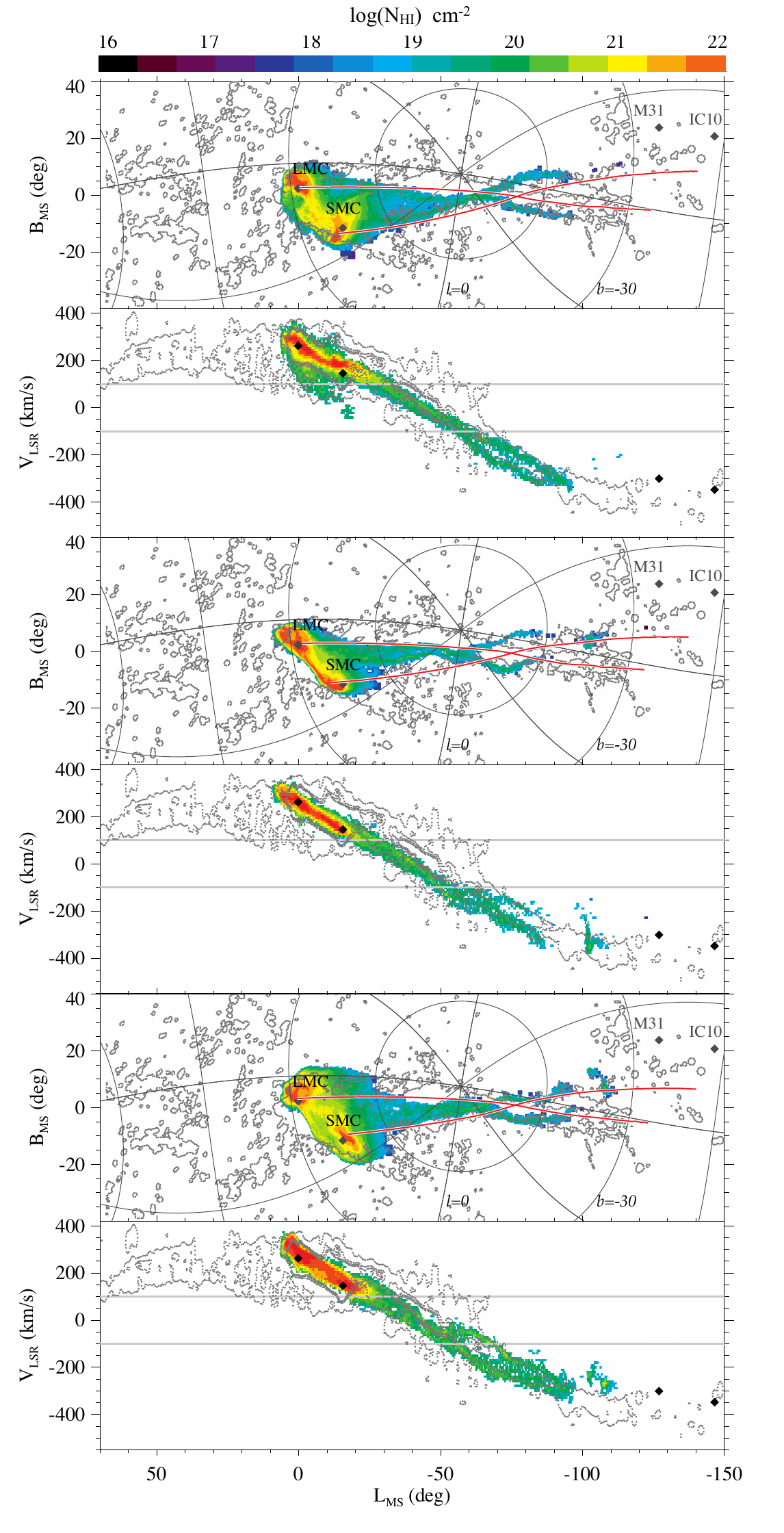}
\caption{Models of the Magellanic Stream N(H I) distribution in the ($L_{MS}, B_{MS}$, top) and ($L_{MS}, V_{LSR}$, bottom) planes as resulting from the 'ram-pressure plus collision' scenario for the MCs, superposed on contours taken from Figure 8 of \citet{Nidever10}. From top to bottom, Model Nos. 385, 395, and 266, which correspond to MC progenitors displayed in Table~\ref{tdg} and MW models shown in Table~\ref{tbmw}. Red lines show the orbits of the LMC and SMC. The top Figure is also available as an mpeg
animation in the electronic edition of the {\it Astrophysical Journal}.}
\label{Fig10_MCs_MS}
\end{figure}

\subsection{Reproducing the MS}
\label{results}
\subsubsection{Comparison with observations}
Figure~\ref{Fig10_MCs_MS} illustrates how the resulting interaction between the MCs and the MW hot gas can form the MS. Initially, the two TDGs are at the location of and with the velocity of the M31 tidal tail. They start to react to the hot gas (appearance of tiny tails), 1 Gyr ago, at $\sim$ 230$\pm$20 kpc from the MW, and then the SMC is passed by the LMC due to the differential ram-pressure effect. They collide with a pericenter of 2.85$\pm$0.4 kpc, 0.25$\pm$0.05 Gyr ago, at 85$\pm$5 kpc and at $L_{MS}$= -80$\degr$ $\pm$5$\degr$ (the $\pm$ sign indicates variations from one model to another). The present-day stage shown in Figure~\ref{Fig10_MCs_MS} reveals two prominent ram-pressure tails trailing behind each MC.

\begin{figure}
\epsscale{1.10}
\plotone{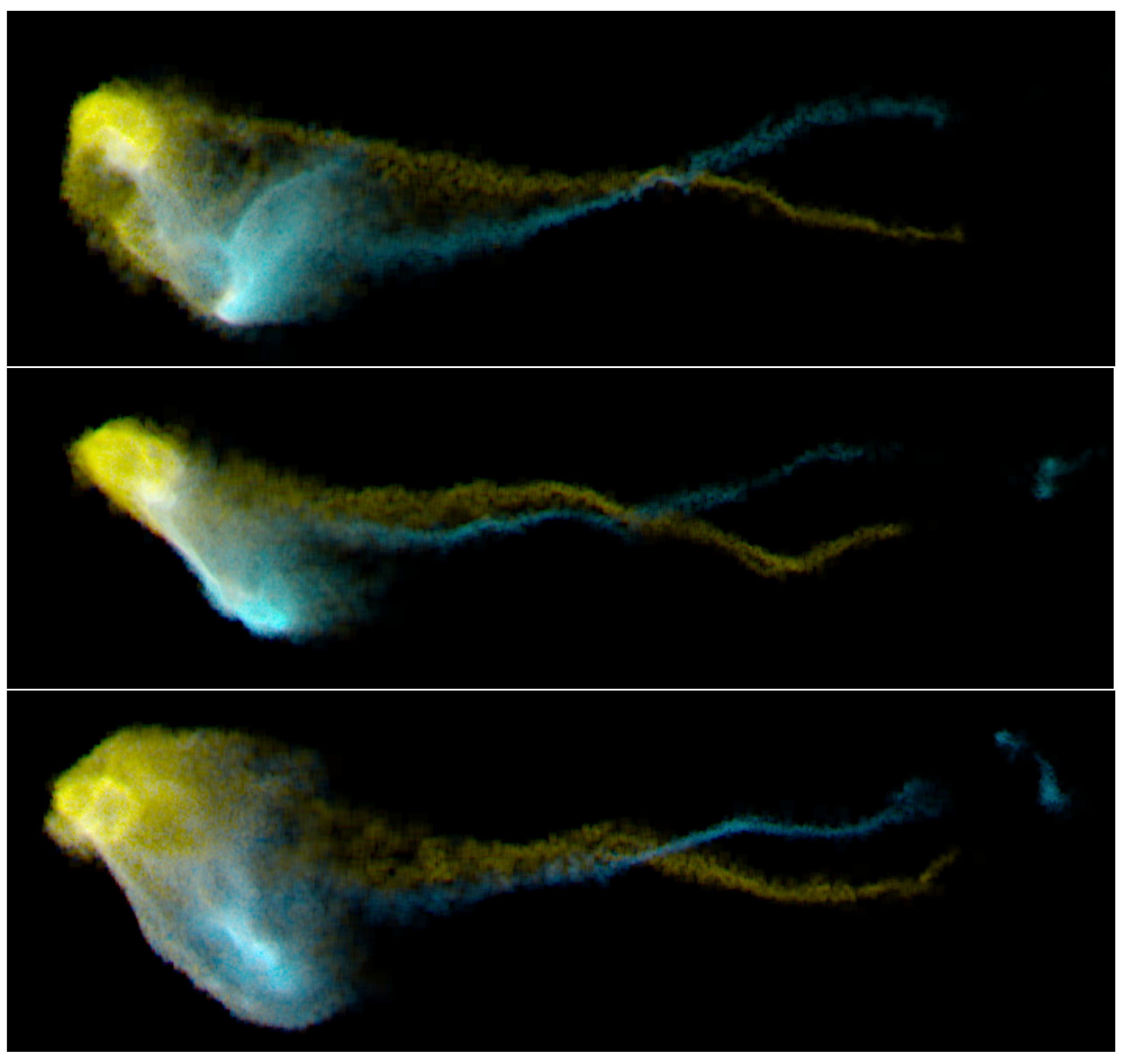}
\caption{Same as Figure~\ref{Fig10_MCs_MS} (top panels), for which we have color-coded particles from the initial LMC and SMC progenitors in yellow and in blue, respectively, to reveal the oscillating filaments and their overlapping.}
\label{Fig11_tails}
\end{figure}

The overall wide shape of the MS from the SGP (or MS-I) to the MCs is reproduced and it is due to the large angle delineated by the two MC trajectories after the collision (see Figure~\ref{Fig5_SGP}, bottom panel). Evidence for this is also provided the two filamentary structures in Figure~\ref{Fig11_tails} for which gas particles from the initial LMC and SMC are marked in yellow and blue colors, respectively. Figures~\ref{Fig10_MCs_MS} and \ref{Fig11_tails} show that from MS-II to MS-IV the twin filaments are almost parallel or intertwisting as observed in Figure~\ref{Fig2_MS_DNA}, and then they bifurcate near the tip end as it is observed near MS-V (see Figure \ref{Fig1_MS}).

 The recent collision also produces a Magellanic Bridge that is as prominent as the observed one, also showing a certain variety of morphologies (see, e.g., Figure~\ref{Fig11_tails}), which depends on the precise collision parameters. Model 395 Bridge is perhaps the closest to the observations, since it reproduces quite well the elongated shape resembling a narrow Bridge, although it is longer than the observed one.

\begin{figure}
\centering
\epsscale{1.10}
\plotone{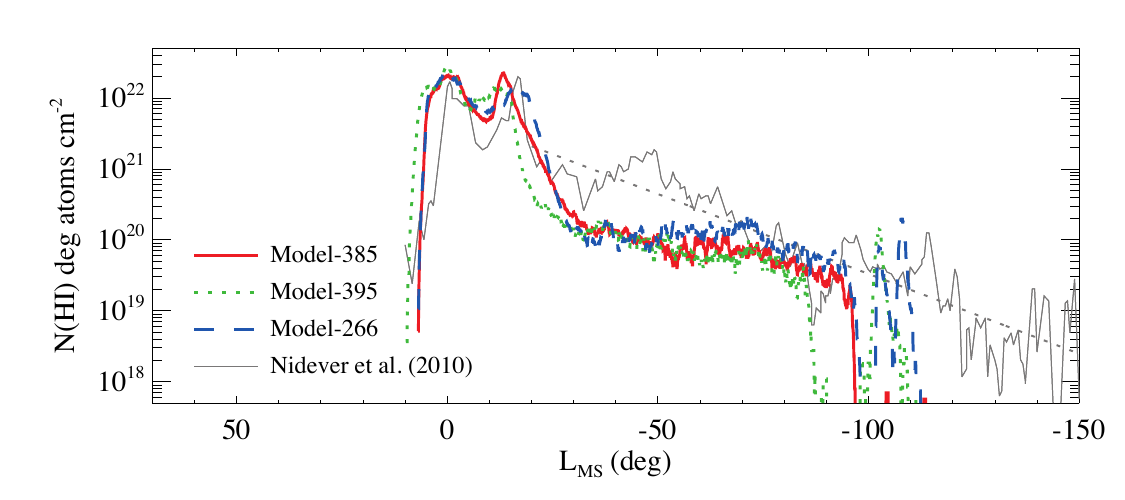}
\caption{Models of the N(H I) profile in the MS,  which are compared to the observed one (solid gray curve and dashed straight line) according to \citet{Nidever10}.
}
\label{Fig12_NHI_modele}
\end{figure}

 Figure~\ref{Fig12_NHI_modele} shows that we recover most of the N(H I) density near the MCs and near the MS tip, while our modeling underpredicts by factor $2-3$ N(H I) near the SGP ($L_{MS}$ from -40$\degr$ to -60$\degr$), where the collision residuals are expected to lie (see Sect.~\ref{residue}). This could be attributed to a difficulty in reproducing the gas expelled during the collision, which seems to be confirmed by the lack of velocity scatter when comparing the modeled MS in the ($L_{MS}, V_{LSR}$) plane to observations (see the underlying contours in Figure~\ref{Fig10_MCs_MS} for $L_{MS}$ ranging from -40$\degr$ to -60$\degr$).

\subsubsection{The role of KH instabilities and survivability of the filaments}
\label{KH_role}

Perhaps an important question is related to the survivability of the filaments, which are the main features of the Stream.  Observations provide us with at least a part of the answer. Indeed we find vortices in the MS filaments from MS-II to MS-IV, and after a careful examination, none from MS-I to the Interface region. This leads us to predict that KH instabilities are more developed in the first region than in the second region, which is likely due to the much larger elapsed time since the material was deposited in the MS-II to MS-IV region. We calculate from Eq. \ref{tKH} that at $L_{MS}$= -40$\degr$ (MS-I), -60$\degr$ (MS-II-III), -80$\degr$ (MS-IV), and -100$\degr$ (MS-VI), $t_{KH}$ values are 213, 142, 100, and 73 Myr, respectively, using $\lambda$= 2 kpc, $V= 300 km s^{-1}$, the hot gas densities from Figure \ref{Fig8_hotgas}, and the cold gas densities from \cite{Nidever10}. Elapsed times are estimated from an examination of our simulations, and their ratios to the KH time are 0.6, 1.3, 2.7, and 6.3\footnote{In principle KH instabilities should have destroyed the MS for such a large ratio according to Figure 25 of \citealt{Hopkins14}. However, MS-VI ($L_{MS}$= -100) is associated to $\rho_{hot}$ $\sim$ 4 $\times 10^{-5}$ $cm^{-3}$, and $Re$ $\sim$  90, i.e., near the critical value that suppresses KH instabilities \citep{Roediger13}.}, respectively. The presence of well defined filaments from the MCs to MS-IV clearly indicates that KH instabilities have not destroyed them, and that to a first approximation, standard SPH codes such as GADGET2 can be used to model them. 

The MS-V-VI regions are possibly more affected due to the relatively large elapsed time ($\sim$ 470 Myr), and indeed the observed filaments are rather dislocated there (see, e.g., Figure \ref{Fig1_MS}). Then our simulations could have lost part of the mixing especially at the MS tip, which could be responsible for our predicted N(H I) profile being flatter than that observed (see Figure \ref{Fig12_NHI_modele}). The above calculations confirm that our modeling has essentially captured the filamentary structure as shown in Figure \ref{Fig11_tails} and revealed by observations. 

\begin{figure}
\epsscale{1.2}
\plotone{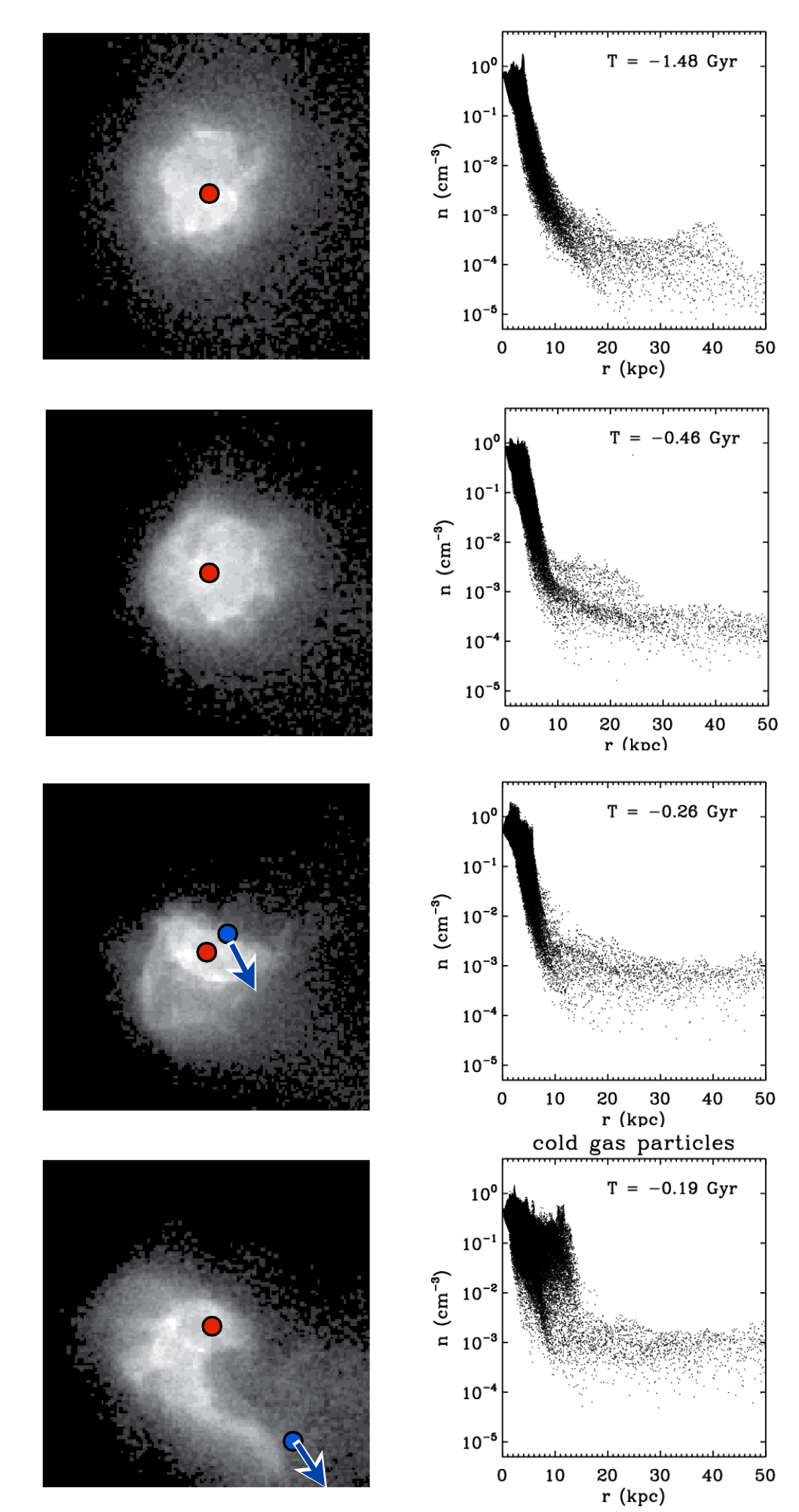}
\caption{Snapshots ($19.8\times19.8 kpc^{2}$) of Model 385 of TDG-A1 gas particles assuming the gas disk is inclined at 29$\degr$ as for the observed LMC \citep{Staveley-Smith03}. On the left, it shows the (R.A., decl.) plane with red and green dots representing the LMC and SMC, respectively, and on the right the density profile. The disk (initially extracted within 20 kpc) at T = -1.48 Gyr shows the supernova imprints (as it is observed in the LMC) expelling gas up to 40 kpc and the gas is distributed almost symmetrically since it is unaffected by the MW hot gas (at 385 kpc and $\rho_{hot}$= 2.4 $\times 10^{-6}$ $cm^{-3}$).  At T= -0.48 Gyr (138 kpc and $\rho_{hot}$= 6.8 $\times 10^{-5}$ $cm^{-3}$) the gas outer disk has significantly shrunk due to ram pressure, especially in the direction of the MW center (left of the image). The pericenter  of the LMC-SMC collision occurs at T = -0.26 Gyr distorting the LMC disk, and is responsible for expelling toward the SMC a large amount of gas material at T = -0.19 Gyr. This Figure is also available as an mpeg animation in the electronic edition of the {\it Astrophysical Journal}.}
\label{Fig13_feddback}
\end{figure}

\subsubsection{The Interplay between Feedback and Ram pressure}
Figure~\ref{Fig13_feddback} shows that feedback helps considerably in expelling gas at the LMC outskirts, which is then subject to ram pressure to form the MS. Such a mechanism actually occurs in the LMC, and \cite{Nidever08} discovered that the LMC filament originates from the star-forming region 30 Doradus. We find that the shape of the Interface region is very sensitive to the precise amount of feedback, as it is illustrated in Figure~\ref{Fig10_MCs_MS} (compare Model 266 with Models 385 and 395, for which feedback changes from 5 to 4 times the median value of \citealt{Cox06}, respectively). 

The role of feedback is strongly supported by H I observations of the LMC \citep{Kim98}, revealing the complex system of filamentary structures, combined with numerous H I holes and shells, the formation mechanism of which is well illustrated in Figure~\ref{Fig13_feddback} (see the video in the online journal). We further notice that overall structure of the LMC is stable even against high values of feedback, and that the collision between the MCs may also play a role in extracting the gas from the LMC disk (see the two bottom panels in Figure~\ref{Fig13_feddback}).

\subsubsection{The possible origin of the huge amount of MS ionized gas}
\label{ionization}
We estimate the ionized gas beyond 30 kpc from the MCs to be 0.44, 0.64, and 0.7 $\times 10^{8}$ $M_{\odot}$ for Models 266, 385, and 395, respectively, to be compared to $\sim$ 2 $\times 10^{9}$ $M_{\odot}$ found by \cite{Fox14}. Such a considerable discrepancy could be due to an improper treatment of either KH instabilities or of the gas expelled during the MC collisions. 

To verify how important the role of KH instabilities is, we have implemented in GADGET an alternative SPH implementation (rpSPH, \citealt{Abel11}), which was designed to better account for them. When compared to our former runs, it increases the ionized gas mass associated with the Stream by only $\sim$ 10\%. 

As shown by Figure~\ref{Fig10_MCs_MS} and Figure~\ref{Fig12_NHI_modele}, the observed gas distribution surrounding the MCs and in the Interface differs from that in our simulations. Within 10 kpc from the TDG-A and B, the neutral gas accounts for $\sim$ 3 $\times 10^{9}$ $M_{\odot}$ instead of 1.1 $\times 10^{9}$ $M_{\odot}$ from \cite{Putman03} in MCs, Bridge, and Interface, together (see also Figure~\ref{Fig12_NHI_modele}). There is perhaps a sufficiently large amount of gas in our modeling to explain the observed one, but it is not distributed as in the observations. 

Based on our analysis of the observations in Sect. \ref{residue}, we conjecture that most of the gas has been ionized during and after the collision. To verify this would require us to capture the overall shock process that could form small gas clumps. The masses of HVCs like those revealed in large numbers in Figure \ref{Fig6_shock} are from 140 to $10^{5} M_{\odot}$ with an average of 1.4 $\times 10^{4} M_{\odot}$ \citep{Putman02}, i.e., even smaller than that of a single gas particle in our initial TDGs. This could explain why our modeling does not capture the bow shock or the large scatter of velocities from $L_{MS}$ = -40$\degr$ to -60$\degr$ (see velocity panels in Figure \ref{Fig10_MCs_MS}).

Further modeling with much higher resolution will be done in a future study. To improve the treatment of fluid mixing instabilities, \cite{Hopkins13} introduced a Lagrangian formulation allowing better performance in treating KH instabilities and shocks. The above conditions have to be used to verify: (1) how the material is expelled into the hot gas and how it is affected by KH instabilities and can reproduce the gigantic shock described in Figure \ref{Fig6_shock};  (2) the precise  encounter between the Clouds and to which conditions it may lead to a lagging structure like the bright spot (see Sect.~\ref{SGP} and Figure \ref{Fig5_SGP}); although the SMC gas disk is always destroyed in our simulations, we are led to conjecture that specific conditions of the encounter may have caused the expulsion of part of it within the MS; and (3) the precise filament morphologies with different vortex structures, and whether they can be caused by different metal abundances (see Sect. \ref{vonKarman} and Figure \ref{Fig2_MS_DNA}).

\subsection{Consistency between TDGs and the Cloud progenitors}
\subsubsection{Baryonic mass and characteristic radius}
\label{MCmasses}

Table~\ref{tdg} gives the stellar and gas content of TDG-A (LMC) and B (SMC). The gas content of TDG-A can be compared to LMC values from \citet{Staveley-Smith03}, who found 0.38 $\times 10^{9}$ $M_{\odot}$ within the disk (r $<$ 3.5kpc), i.e., 0.52 $\times 10^{9}$ $M_{\odot}$ when accounting for helium. The latter value is in relatively good agreement with values in Table~\ref{tdg} (from 0.49 to 0.72 $\times 10^{9}$ $M_{\odot}$).  \cite{Stanimirovic99} found a SMC gas mass of 0.6 $\times 10^{9}$ $M_{\odot}$ after helium correction over an area of $\sim$ 24.5 $kpc^{2}$, which is higher than our modeled TDG-B ($0.3-0.48 \times 10^{9}$ $M_{\odot}$ in a 1.5 times larger area), but both observations and simulations show very irregular distributions that complicates the comparison. While Table~\ref{tdg} shows that the gas contents of both LMC and SMC are approximately reproduced, their stellar contents are consistent with values ($L_V$= 1.5 and 0.46 $\times 10^{9}$ $L_{\odot}$ for the LMC and SMC respectively) from \cite{McConnachie12} assuming that $M/L_V$ may vary from 1 (TDG-A1, B1) to 2.2 (TDG-A2, B2), the latter value being adopted by \cite{Harris09}. 

Table~\ref{tdg} provides values of stellar radius for TDG-A that are about two thirds of the observed ones, assuming that disk scale length of the LMC is 1.4 kpc \citep{Weinberg01}, and $r_{half}$ $\sim$ 2.35 kpc (or 2.2 kpc if adopting the \citealt{vanderMarel06} value). The radius delimiting the LMC gas disk is 3.5 kpc (quoted from \citealt{Staveley-Smith03}), a value corresponding to (and consistent with) the gas cut-off radii found in Table~\ref{tdg}, but for Model 395 for which the gaseous disk has been too severely shrunk by the collision. Then we conclude that MC baryonic masses, gas fractions, and radii are reproduced to first order by our simulations.

Admittedly, even if their baryonic mass and size properties are comparable, our TDG-A and B are not precisely the LMC and SMC, and further simulations are required to reproduce them accurately. On the other hand, our results suggest that for each mass distribution of initial TDGs, one is able to derive a MW hot gas profile that allows one to reproduce the Stream and its double filamentary structure. For example, the initially gas-rich TDG-A2 and B2 (Model 266) require a lower hot gas density (MW122) than what is needed for less gas-rich  TDG-A1 and B1 (Models 285 and 295 associated to MW 165 and 166, respectively, see Figure~\ref{Fig8_hotgas}). Similarly, the passage of massive, dark matter-dominated galaxies would require a denser hot gas\footnote{The argument of \cite{Grcevich09} (see also their Equation 1) is based on the gas stripping of $<$ 270 kpc dSphs assuming that their masses are supported by their velocity dispersions. To strip the gas from their progenitors requires a far larger hot gas density than in this paper or in \cite{Yang14}, who assumed TDGs as dSph progenitors and found their kinematics to be consistent with an expansion of their stellar content.} ($\rho_{hot}$= $(2-3) \times 10^{-4}$ $cm^{-3}$ at 100-200 kpc, see the points from \citealt{Grcevich09} in Figure~\ref{Fig8_hotgas}) than what we use in this paper or what is found by \cite{Kalberla06} in the vicinity of the MS.

Then it is plausible that the real MCs may have formed two ram-pressure tails after entering into the MW halo hot gas. 

\subsubsection{Kinematics}
\label{LMCrot}
\begin{figure*}
\centering
\epsscale{1.0}
\plotone{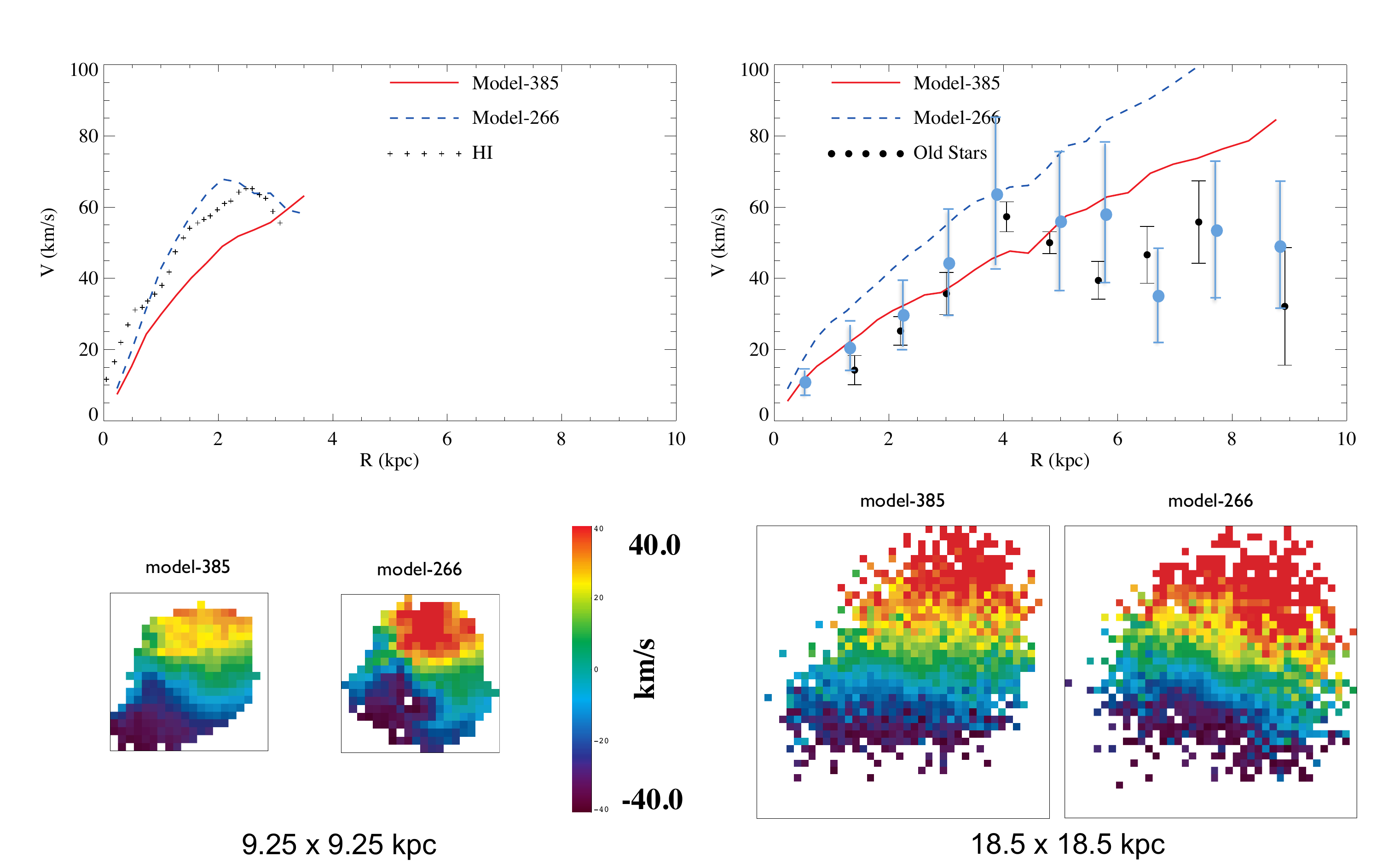}
\caption{{\it Top:} rotation curves derived from the H I gas (left, small crosses from \citealt{Kim98}) and for old stars (right, small black and large blue dots from \citealt{vanderMarel02,vanderMarel14}, respectively). Error bars for the points of \citet{vanderMarel14} account for the inclination-related uncertainties as described in their paper, which have been quadratically added to the shot noise quoted in their Table 4. Curves show the result of the KINEMETRY analysis (see, e.g., \citealt{Krajnovic06}) applied on Models 266 and 385 with a limiting radius of 3.5 and 10 kpc for the gas and $>$ 2Gyr old stars, respectively. {\it Bottom}: from left to right, the velocity fields derived for Models 266, 385 for the gas (for i= 39$\degr$ and 29$\degr$) and for old stars (for i= 35$\degr$ and 44$\degr$), respectively.}
\label{Fig14_LMCrc3}
\end{figure*}

Figure~\ref{Fig14_LMCrc3} shows that the rotation curves of Models 266 and 385 are rather consistent with that of the LMC, for both gas and old stars within their disk radii, 3.5 and 5.5 kpc, respectively. Comparison with H I observations of \cite{Kim98} is quite easy since they have used the the standard tilted ring algorithm ROCUR, which is quite equivalent to our use of the algorithm KINEMETRY that fits the best ellipse along the profile of the galaxy, decomposing the velocity field in harmonic expansion (see, e.g., a detailed comparison in Sect. 4.3 of \citealt{Corbelli14} and their Figure 9). We have fixed the inclination of the disk from gas and stellar contours, although our results are almost unchanged by retrieving the inclination from the KINEMETRY software. 

Comparison to the old star observations \citep{vanderMarel02,vanderMarel14} is slightly complicated by the fact that they performed a $\chi^{2}$ fit for modeling the LMC velocity field, assuming it is a purely rotating disk. Thus we are cautious that some discrepancies may occur because of the different adopted methodologies, although this is limited by the fact that the rotation is the dominant component for both the LMC and simulated TDG-A, at least within the optical disk (R $\le$ 5.5 kpc). Within the latter, our modeling is quite similar to the observations, while beyond it, modeling reveals motions that are in excess of the observed ones.

In fact the observed LMC is with a morphological Position Angle (PA) that is $\sim$ $40\degr-50\degr$ off from the kinematic PA \citep{vanderMarel14}, a property shared by our simulated TDG-A and that we attribute to the recent collision. Such a discrepancy usually excludes a galaxy from being classified as a rotational disk (see, e.g., \citealt{Garrido02,Epinat10,Garcia-Lorenzo14}). The interaction between LMC and SMC shown by the Bridge could explain this as well as why the apparent rotation at the LMC outskirts (see Figure~\ref{Fig14_LMCrc3}) exceeds expectations from its baryonic mass content alone.

Then it results that our modeling is consistent to first order with the LMC kinematics.

Modeling the LMC as a TDG remnant, however, is non-trivial (see \citealt{Hammer13}) even if they possess similarly thick disks with $V/\sigma$= 2.9 and 2.7 for LMC carbon stars \citep{vanderMarel02} and from our modeling, respectively.  Perhaps surprisingly, the LMC stellar mass is not too large to be linked to a TDG, because $\sim$ 5\% of the TDGs observed by \cite{Kaviraj12} have stellar masses comparable to or larger than that. Figure 6 of \cite{Kaviraj12} compares the stellar mass ratio between the parent galaxy and the TDG, which would be approximately 1/55 when comparing LMC to M31. If associated with  an ancient gas-rich merger that formed M31, the LMC would be within the 25\% of TDGs having a mass ratio similar to or larger than 1/55 of their parent galaxy.

\subsubsection{LMC PMs}
\label{PMs}

\begin{figure}
\epsscale{1}
\plotone{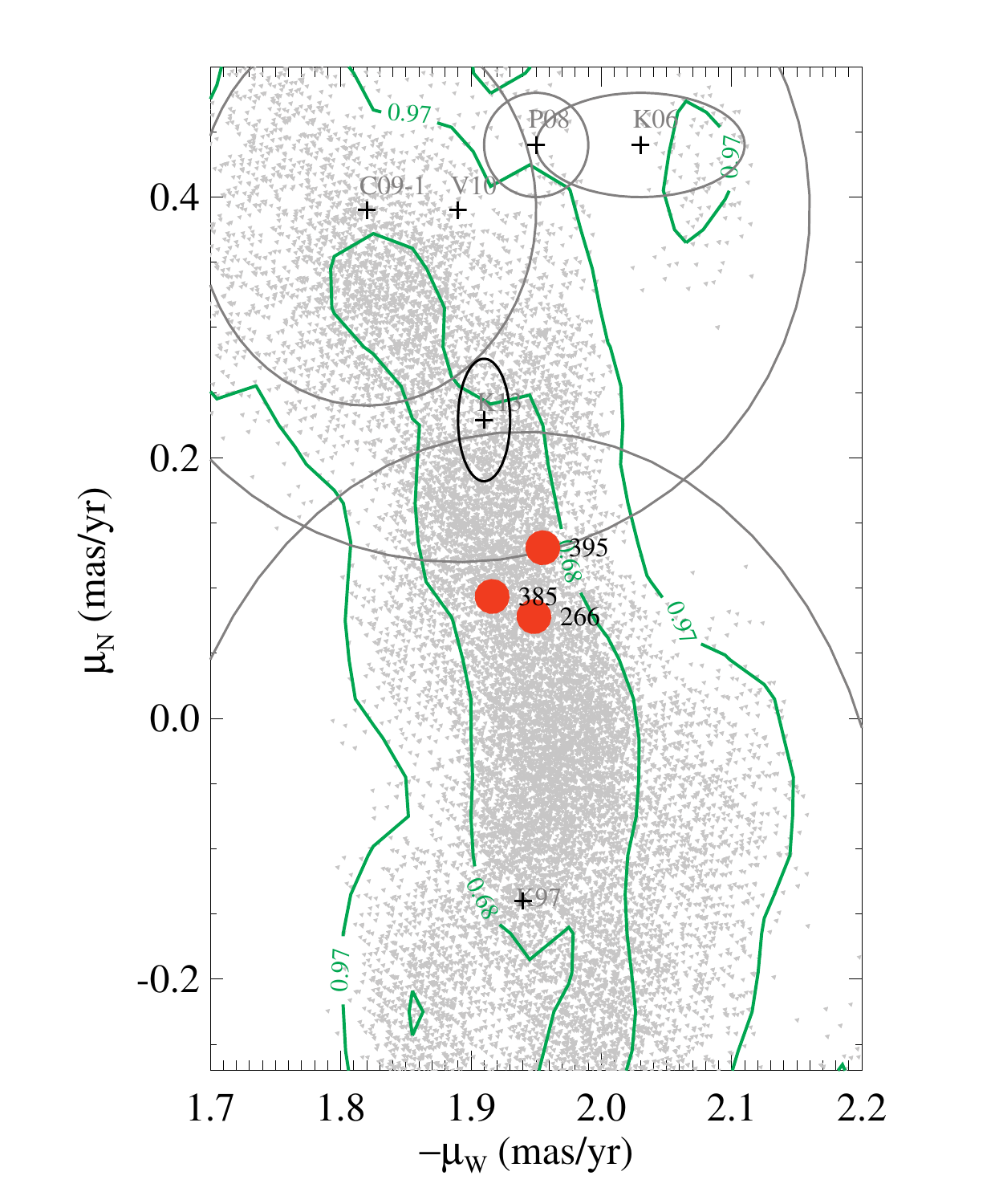}
\caption{LMC proper motions and uncertainties (ellipses in this zoomed version of Figure 10 of \citealt{Vieira10}, an estimate that is centered on V10),  determined by K97 \citep{Kroupa97}, K06 \citep{Kallivayalil06}, P08 \citep{Piatek08}, C09-1 \citep{Costa09} together with the most accurate value from \citet[]{Kallivayalil13}[K13] and its $1\sigma$ ellipse. Motions of the mass centers of our models are labeled by their number (red filled circles). The contours show the probabilities when assuming the final LMC extracted from these models, about 200 Myr after its collision with the SMC (see text).
}
\label{Fig15_PMs}
\end{figure}

The PMs measured by \cite{Kallivayalil13} lead to orbits that significantly deviate from the MS, and this could be a major argument against a pure ram-pressure model (see, e.g., \citealt{Besla12}).   
On the other hand our modeling is driven by the equally robust observations of the filamentary nature of the MS as well as of the relics of the recent collision between SMC and LMC. The latter event is certainly to be considered when extracting the measurements of star motions in Hubble Space Telescope (HST) fields to derive the overall LMC PM.  At first sight, our three models are consistent within 2-3 $\sigma$ with \cite{Kallivayalil13} PMs (compare their locations to the 1$\sigma$ ellipse surrounding the K13 point in Figure~\ref{Fig15_PMs}). We have, however, to account for the fact that after the collision, the LMC is not an ideal rotating disk as it was assumed by \cite{vanderMarel14}.

To account for these effects we have:
\begin{enumerate}
\item reproduced the calculations of \cite{vanderMarel14} of PMs and associated uncertainties (see their Table 1), providing very similar values in the context of their adopted pure rotation for the LMC;
\item replaced the pure rotation model by the final TDG-A extracted from the three models that reproduce the MS fairly well (see Figure~\ref{Fig10_MCs_MS}) and the LMC rotation curve rather well (see Figure~\ref{Fig14_LMCrc3}).
Their disks are indeed affected by the collision, and their kinematic axes deviate significantly from their PAs as for the observed LMC. 
\item To verify the consistency of our modeling hypothesis with the PMs of \cite{Kallivayalil13}, we have performed Monte Carlo simulations to randomize the systematic errors related to the non-circularity of the LMC disk. Figure~\ref{Fig15_PMs} shows that the PMs of \cite{Kallivayalil13} are consistent with the ram-pressure plus collision model within $\sim$ 68\% confidence level (equivalent to $1\sigma$). 
 \end{enumerate}

Interestingly the systematic errors are elongated in the N-S direction (affecting $\mu_N$), which could be attributed to the fact that only one of the ten three-epoch HST fields lies to the south of the LMC. Variations of the  motion of the LMC  center of mass (COM) are mostly linked to variations of $\mu_W$ in order to match the MS location and velocity profile. We have analyzed the three models presented in this paper as well as all other simulations that we made during the optimization of the MS (see Sect.~\ref{calcul_orbit}), under the condition that they populate the $\ge$ 68\% level area in Figure~\ref{Fig15_PMs}. We verified that we can generate the MS with velocities ranging from $\sim$ 250 to $\sim$ 350 km$s^{-1}$, but not below or above these values. Thus the characteristic times such as studied by \cite{Bland-Hawthorn07}, \cite{Tepper-Garcia15} and in this paper do not seem to be very much affected by such systematics.

A full description of the above analysis and calculations can be found in the Appendix A, showing that when accounting for the recent Cloud collision, the ram-pressure model can be reconciled with PM measurements.

\subsection{Toward the complete reproduction of the Magellanic System}
\label{discussion}

\begin{figure*}
\centering
\epsscale{1.2}
\plotone{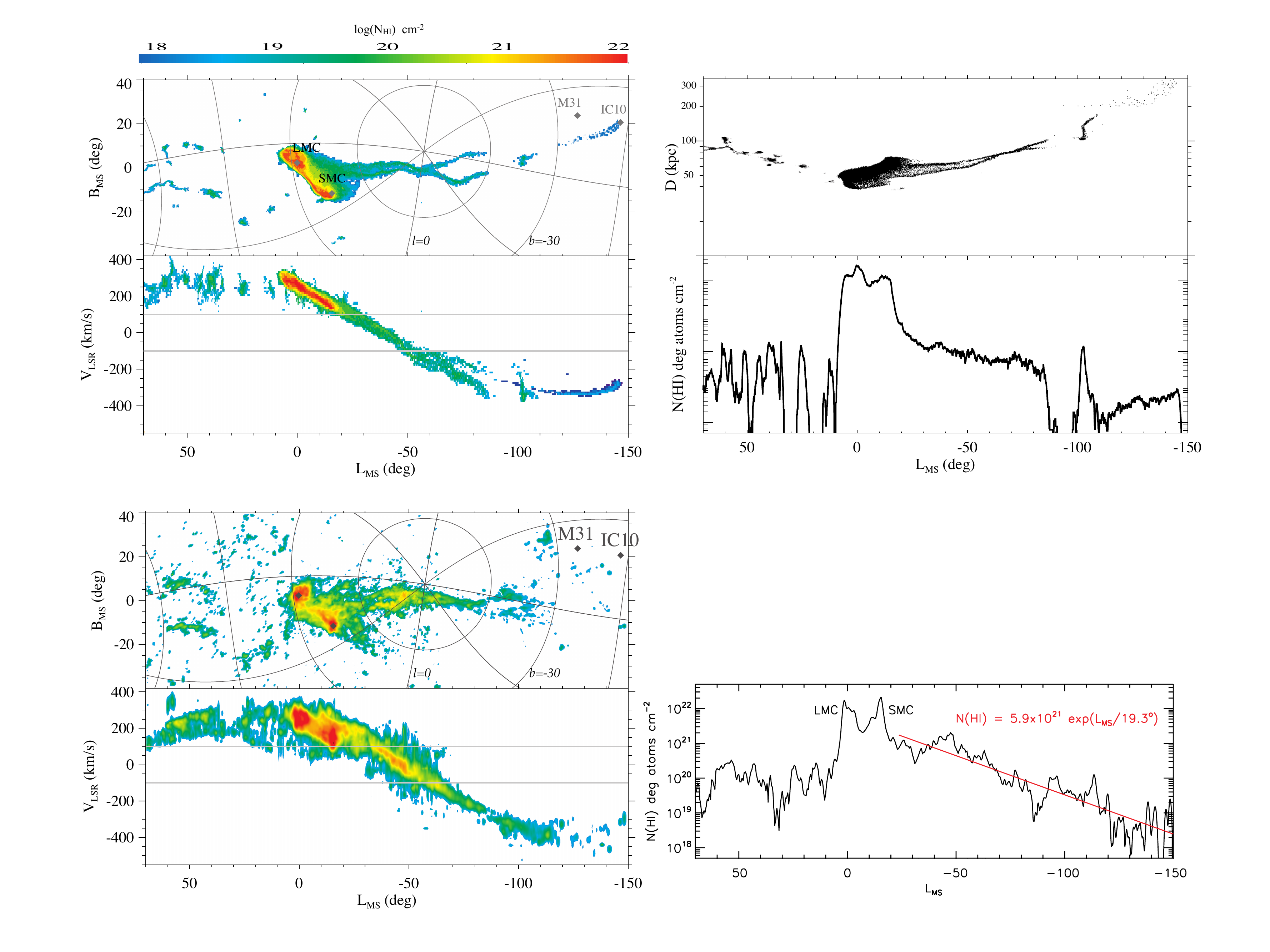}
\caption{{\it Top four panels}: same as Figure~\ref{Fig10_MCs_MS} and Figure~\ref{Fig12_NHI_modele} for Model 395 with the addition of a top-right panel showing the MS distance versus $L_{MS}$ providing values quite similar to that of \cite{Guglielmo14} and much smaller than those of \cite{Besla12}, especially their Model 2. The main difference with Figure~\ref{Fig10_MCs_MS} arises from the fact that  the simulation of the MS is complemented by (1) that of the three previously stripped dwarfs, progenitors of MW dSphs, the gas from which has been deposited onto the Leading Arm (see left side in each panel), and (2) that of the projected M31 tidal tail pointing to IC 10 (see right side in each panel). {\it Bottom three panels}: it shows the observations from Figures 8 and 10 of \citet{Nidever10} for comparison. Our low-resolution simulations do not retrieve small gas clumps, extended low N(H I) features near the MCs, or the large velocity scatter near $L_{MS}$=-50$\degr$ (see text).
}
\label{Fig16_full}
\end{figure*}

 We are proposing a new interpretation of the whole Magellanic System and its origin (see Sect. \ref{MS_origins}) based on the ram-pressure stripping of progenitors of the Clouds and MW dSphs.

Figure~\ref{Fig16_full} summarizes the full ram-pressure scenario, including the formation of the MS by the MCs and that of the LA by former passages of dSph progenitors, which have entered the MW halo $410-500$ Myr earlier than the progenitors of the Clouds. The TDGs originate from a past merger in M31, the tidal tail relics of which coincide with the S0 filament found by \cite{Nidever13} in both space and velocity. 
Figure~\ref{Fig16_full} decomposes the Magellanic System into three main Stream elements (the LA, the MS double filamentary structure with imprints of the recent MC collision, and S0 representing the M31 tidal tail), and within the single frame of an interaction of the MW with the TDG-populated, M31 tidal tail. The fact that several dSphs may have deposited gas near the LA region helps to reproduce the multiple structures found by \cite{Venzmer12}, while their oscillating behavior is warranted by the hot medium in which they are bathed. Such a behavior was indeed noticed by \cite{Yang14} with a different  density profile for the halo hot gas and it is shared by all the models studied in this paper. 

The 'ram-pressure plus collision' scenario imprinted in the Stream (see  Figure~\ref{Fig7_MS}) and modeled in Figure~\ref{Fig16_full} provides a comprehensive, physically driven interpretation of its formation. It reproduces quite well the morphological and kinematic behavior of the whole Magellanic System, as well as the overall N(H I) profile along $L_{MS}$ (from \citealt{Nidever10}, see Figure~\ref{Fig16_full}), over its 230$\degr$ extent.

\section{Conclusions}
We have analyzed Magellanic System observations, using the best observations ever made of the Southern Hemisphere H I, i.e., the GASS survey from \cite{McClure09} and \cite{Kalberla10}. Both the MS and LA are dominated by turbulence, explaining well their oscillating behaviors. The presence of different types of vortices allows us to decompose the Magellanic Stream (MS) into two filaments, one lagging behind each MC, as expected for ram-pressure tails. The distribution of the vortices follows expectations for two, transonic, von Karman vortex streets (see Figure~\ref{Fig2_MS_DNA}) that are intertwisting to form a DNA-like behavior from MS-II to MS-IV as described by \cite{Putman03}.

Examining the data near the SGP (between MS-II and MS-I), we have used the remarkable spectral resolution of GASS to disentangle the MS emission from the MW gas disk contamination. We have identified at the MS peak emission a probable residue of the MC collision in the form of a high-density blob at l = 306$\degr$ and b = -79$\degr$ in the MS, surrounded by a huge concentration of HVCs along a line almost perpendicular to the MS (see Figure~\ref{Fig6_shock}). This resembles the signature of a gigantic bow shock due to clouds ejected during the collision and heated by the hot gas, which is furthermore supported by the large velocity scatter of these anomalous HVCs (see also \citealt{Putman03}). Past events are imprinted into the MS gas distribution, leading to an accurate description of the MC progenitor orbits. From MS-II to MS-IV, filaments were nearly parallel until the collision residue, where they abruptly change to a wide angle as seen in Figure~\ref{Fig5_SGP} (bottom panel). This provides further evidence of a radical change of orbits especially that of the SMC, $200-300$ Myr ago. 

Following a scenario guided by the ram-pressure exerted by the MW hot gas on gas-rich, infalling dwarfs, we modeled their collision and gas exchanges. Progenitors are launched using the GADGET2 SPH code, which describes how the MS is formed by the interplay between feedback from supernovas that eject sufficiently large amounts of gas to feed ram pressure from the hot gas of the MW halo. Such a physical mechanism is observed in the present-day LMC \citep{Nidever08}, and our modeling reproduces the observations at a level never before reached (see Figure~\ref{Fig16_full}). For the first time, the whole Magellanic System is interpreted and successfully understood in a single physical model, including its overall morphology, and its velocity and H I column density profiles along $L_{MS}$. It assumes that the consecutive passages of the MCs and of dSph progenitors have formed the MS and the LA, respectively. We notice that such an alignment of trajectories along the MS between those dwarfs is expected if they are TDGs that all belong to the MW plane of satellites (e.g., VPOS, \citealt{Pawlowski11}). The remarkable success of the "ram-pressure stripping plus MC collision" in reproducing most features of the Magellanic System makes it as the most serious contender for explaining its formation.

 Reproducing the detailed morphology and kinematics of the gigantic Magellanic System still requires important improvements. 
 First, KH instabilities are not well handled by SPH codes such as GADGET2 (see Sect. \ref{KH_role}), and by using GIZMO \citep{Hopkins14} one could also reproduce many more details of the Stream including the associated vortices. Second, it is necessary to use high-resolution simulations of the precise collision between MCs to reproduce its residuals in the form of the numerous anomalous HVCs and the bright spot near the SGP, which we conjecture could be a specific residue of the destruction of the SMC H I disk under certain orbital conditions. Such future studies may account for far more constraints from the detailed structures of the Clouds, the Bridge, and the Interface area, as well as providing an explanation for the huge amount and area of ionized gas, which we suspect could be mainly linked to the gas expelled during and after the collision.

\acknowledgments

We thank the China-France International Associated Laboratory Origins
supported by the Chinese Academy of Sciences, the National
Astronomical Observatory of China, the Centre National de la Recherche
Scientifique and the Observatoire de Paris. We are very grateful to David
Nidever for his kind sending of MS data in numerical format that have been used
for this paper. We are also considerably indebted to the GASS team \citep{McClure09,Kalberla10} for making publicly available such excellent data. 
We thank Pavel Kroupa and Marcel Pawlowski for their very useful remarks. We are grateful to the referee for helping us in structuring the paper and adding essential and useful references. 
This work was granted access
to the HPC resources of MesoPSL financed by the Region
Ile de France and the project Equip@Meso (reference ANR-10-EQPX-29-01) of the programme Investissements d'Avenir
supervised by the Agence Nationale pour la Recherche.






\appendix

\section{Calculations of PMs}
The most accurate determination of the LMC PM to date is that of
\cite{vanderMarel14} and \cite{Kallivayalil13} (see Figure
\ref{Fig15_PMs}). They used HST stellar fields to determine PMs in
different regions, mostly in the northern region of the LMC.
\cite{vanderMarel02} showed that the LMC kinematics can be modeled as
the superposition of two main components, i.e., the LMC internal
kinematics (modeled as a perfect rotating thin disk) and the systemic
motion of the COM. This second term can be further
decomposed into the constant PM of the LMC COM (which is what we are
looking for) and a spatially varying component that depends on the
position of the LMC relative to the observer. As such, the LMC COM PM
is determined as the kinematic contribution left once the best-fit
rotating disk model relative to the observed stellar PMs has been
subtracted, along with the spatially varying term of the systematic
component. This uncertainty on the LMC kinematics could add a systematic error linked to the choice of the model.

Such a source of error might be non-negligible, because of (1)  the recent collision between LMC and SMC, ( 2)  the fact that there is a  50$\degr$ offset between kinematic (from both stars and H I) and photometric PAs as described by \citet{vanderMarel14}, ( 3) the H I, bar, stellar, and PM dynamical centers that are significantly offset from one another ( see discussion in \citealt{vanderMarel14}).

Estimating the systematic uncertainty associated with the modeling of
the internal kinematics of the LMC is difficult because one does not
know precisely what is the exact dynamical state of the LMC. Here, we followed a
pragmatic approach and used the results of the MS modeling shown in
Figure \ref{Fig10_MCs_MS}. These models succeed in reproducing the prominent features of the MS and
the general properties of the MCs (see Sect.~\ref{results} and  \ref{MCmasses}).They therefore have the advantage of providing an internally
consistent estimate of the impact of the dynamical state of the LMC on
the determination of its PM.

We therefore used the three MS models presented in Figure
\ref{Fig15_PMs} from which snapshots around the LMC position at
the present day were extracted. These snapshots need to be
reoriented and rescaled to match how the real LMC is observed on the
sky relative to us and therefore to account for exactly the same
projection effects. In practice, for simplicity, we rotated the
positions of the fields of \cite{vanderMarel14} so that the relative
positions of these fields compared to the model kinematic PA matches
the LMC observations. The models were then rotated in steps of 5$\degr$
along the rotation plane and then re-projected on the plane of the sky
in order to sample different possible morpho-kinematic misalignments.
From each snapshot, the mean PMs of the stellar particles located in
the observational fields were determined. These were used to
determine the best-fit rotating disk model, exactly as
\cite{vanderMarel14} did for the analysis of the HST data for the 
LMC. We used exactly the same method as the one used by
\cite{vanderMarel14}, as detailed in \cite{vanderMarel02}. We ensured
that the same kinematic PM fit (including error bars) was obtained
using the observed HST PMs used by \cite{vanderMarel14}. Using this
method, we fitted each of the simulated snapshots to estimate the
possible systematics impacting the estimation of the LMC PMs when assuming
a perfect rotating disk.

To compare with the observed LMC PM derived by \cite{vanderMarel14},
the total uncertainty on the model COM PMs must take into account both
random and systematic uncertainties. To combine both effects, we
considered each of the ~200 simulated and representative LMC described
above, which were randomized using Monte Carlo simulations assuming
Gaussian distributions with standard deviations as determined by
\cite{vanderMarel14}. The resulting $\sim$ 2$\times10^{5}$ simulations are shown
as gray points in Figure \ref{Fig15_PMs}. We also overlay isodensities
corresponding to the 67 and 96\% percentiles of the 2D density in the
PM plane in bins of 0.025 mas $yr^{-1}$. These would correspond to 1$\sigma$ and 2$\sigma$ if the 2D density were drawn from a Gaussian distribution. This
figure reveals that systematic errors on the COM PMs of LMC-like
systems can have a large and significant impact on the estimated
value, which can be shifted from the true one by several tenths of mas $yr^{-1}$
along both axes. We caution the reader that the isodensities shown in
Figure \ref{Fig15_PMs} should not be considered as real probability
thresholds since, again, we cannot guarantee at this point that the
simulations used are fully representative of the LMC internal
kinematics. Nevertheless, Figure \ref{Fig15_PMs} shows that systematic
effects are large enough to produce the MS models at less than $\sim$
1$\sigma$ (random plus systematic) from the observational estimate of
the LMC PMs by \cite{Kallivayalil13}.




\end{document}